\documentclass[reprint,amsmath,amssymb,aps,prb,eqsecnum]{revtex4-1}
\usepackage[utf8]{inputenc}
\usepackage{physics}
\usepackage{tikz}
\usepackage{graphicx}
\usepackage{color}
\usepackage{url}
\usepackage[colorlinks=true,citecolor=blue,urlcolor=blue]{hyperref}
\let\vec\mathbf

\begin{document}

\title{Magnetic Effects on Topological Chiral Channels in Bilayer Graphene}
\author{Aaron Winn}
\author{John Maier}
\author{Andrew Jiao}
\author{Jeffrey C. Y. Teo}
\affiliation{Department of Physics, University of Virginia, Charlottesville, VA, 22904}
\date{\today}

\begin{abstract}
We study the effect of a magnetic field on topological chiral channels of bilayer graphene at electric domain walls. The persistence of chiral edge states is attributed to the difference in valley Chern number in the regions of opposite electric field. We explore the regime of large electric and magnetic fields perpendicular to the lattice. The magnetic field shifts the channel away from our electric interface in a way that is inconsistent with the semiclassical expectation from the Lorentz force. Moreover, the magnetic field causes an imbalanced layer occupation preference to the chiral channels. These behaviors admit analytic solutions in the limits that either the electric or the magnetic field dominates. We numerically show in the general case that the system can be well-approximated as a weighted sum of the two limits.
\end{abstract}

\maketitle

\section{Introduction}\label{sec:intro}
In recent years, there has been considerable interest in the electronic properties of bilayer graphene~\cite{McCann07,NetoGuineaPeresNovoselovGeim09,McCann13,ROZHKOV20161} including a quadratic dispersion with Berry phase of $2\pi$ and an unusual quantum Hall effect~\cite{Kovoselov06,FalKo08}. The inequivalence of the $K$ and $K'$ Dirac points in the Brillouin zone leads to a valley degree of freedom, which can play a similar role as electron spins and may be manipulated in ``valleytronic" devices~\cite{RycerzTworzydloBeenakker07,Behnia12,AngYangZhangZhongshuiAng17}. New techniques have also made it possible to produce large flakes of Bernal (AB-stacked) graphene on the order of $10^{-4}$ m,~\cite{Ji17} motivating the experimental study of low-energy electron transport. 

Notably, a tunable bandgap has been realized both theoretically as a result of layer-asymmetry between the on-site energies~\cite{Min07} and experimentally by applying an electric field perpendicular to the layers~\cite{Zhang09}. In pure samples where inter-valley mixing is suppressed~\cite{MorozovNovoselovKatsnelsonSchedinPonomarenkoJiangGeim06,McCannKechedzhiFalkoSuzuuraAndoAltshuler06,MorpurgoGuinea06}, a non-local valley symmetry emerges and the insulating electronic system carries a non-trivial valley Chern number~\cite{ZhangMacDonaldMele13}. This symmetry-protected topological nature~\cite{VaeziLiangNgaiYangKim13} of the ground state is revealed by a one-dimensional highway of electrons along an electric domain wall~\cite{MartinBlanterMorpurgo08,QiaoJungNiuMacDonald11,JungZhangQiaoMacDonald11,ZareniaPereiraFariasPeeters11}, across which the interlayer potential difference, and consequently the valley Chern number, change sign. The electronic highway consists of four counter-propagating pairs of chiral channels, where electrons are laterally confined but delocalized along the line interface. The forward and backward propagating directions are locked with the two valley index, so that all electron modes of the same valley species propagate in the same direction. Inter-valley scatterings are forbidden by the valley symmetry and lattice momentum conservation along generic line interfaces~\footnote{Except along commensurate ones, such as the armchair edge, where the valley momentum difference is projected out, ${\bf a}_{\parallel}\cdot(K-K')\in2\pi\mathbb{Z}$ for the primitive parallel lattice vector ${\bf a}_{\parallel}$.}. Moreover, they are chiral anomalous and associate a non-conservative valley current under an interface-parallel electric field. The anomaly is resolved by connecting the high-energy modes to a higher dimensional topological bulk, in this case sandwiched between two valley Chern insulators with opposite topological indices, or allowing the switching between valleys in high-energy. 

These ballistic topological chiral channels have been realized and experimentally observed in bilayer graphene electric domain walls~\cite{Ju15,Li16,Lee17,Li18}. The valley-symmetry-protected channels carry a quantized differential conductance $\sigma_0=dI/dV=4e^2/h$ at zero bias, where the factor of 4 associates to the four counter-propagating pairs of Dirac modes along the interface. This value has been reported in Ref.~\onlinecite{Lee17} using bilayer graphene encapsulated between atomically clean hexagonal boron nitride single crystals, which has been implemented in similar systems to reduce intervalley scatterings~\cite{DeanHone10,KimLee16}. On the other hand, intervalley scatterings, which may be induced by local disorder from uneven substrates and gates among other factors, are non-negligible in other setups. Ref.~\onlinecite{Ju15} reported a domain wall mean free path of $l_0\sim400$ nm, and Ref.~\onlinecite{Li16} inferred a mean free path of $l_0\sim200$ nm. Both are shorter than or comparable to a typical domain wall length $L$, which can range between $200$ nm to 1 $\mu$m. This leads to a significant deviation of the differential conductance from the theoretical quantized value, $\sigma=\sigma_0/(1+L/l_0)$, according to the Landauer–Büttiker formula~\cite{Datta95}. Intervalley scattering can be suppressed in these systems by a uniform perpendicular magnetic field $B$. A $4e^2/h$ approaching differential conductance has been observed in ref.~\onlinecite{Li16} at around $B=8$ T for $L=0.4$ $\mu$m, and similar ballistic transport has been reported in Ref.~\onlinecite{Li18} under similar field strength and length scale. The realization of differential conductance approaching the ballistic limit without a magnetic field in Ref.~\onlinecite{Lee17} suggests the observed conductance cannot be attributed to the quantum Hall effect. 

In addition to enhancing the robustness of the topological chiral channels, the magnetic field provides an external parameter that controls the microscopic properties of the electron modes~\cite{ZareniaPereiraFariasPeeters11,WangRenDengYangJungQiao17}. In particular, we focus on the asymmetric lateral and layer distribution~\cite{Li16} of the electronic wavefunctions due to explicit symmetry breaking by the magnetic field. The wavefunction deviation from the domain wall center position accounts for some of the ballistic transport phenomena observed in a valley valve and electron beam splitter~\cite{Li18}, and introduces new complexities to the critical transport behaviors of bilayer graphene domain wall quantum point contacts~\cite{WiederZhangKane15}. In this paper, we concentrate on the single-body non-interacting understanding and quantitative description of a single domain wall under a large electric potential kink and a strong magnetic field. With potential future applications in micro- or nano-circuit devices in mind, we are interested in the dependence of microscopic quantities, for example the average channel position and lateral localization, on external control parameters, such as Fermi energy, electric and magnetic field. We show that a semiclassical description of these topological channels is inaccurate. For instance, the spatial deviation of the chiral channels under a magnetic field is inconsistent with the prediction using the classical Lorentz force, and can only be explained using Landau level physics.

This paper is organized as follows. In Section \ref{sec:tightbinding}, we review the tight-binding model applied to bilayer graphene in the presence of a uniform external electric field. The Bloch Hamiltonian for bilayer graphene is calculated and linearized about the Dirac points using the k $\cdot$ p approximation. Section \ref{sec:landau} discusses how the Hamiltonian is modified in the presence of a strong magnetic field. Analytic solutions for the energy levels, wavefunctions, average position, localization, and layer preference are found. The emergence of chiral modes at electric domain walls is derived in Section \ref{sec:chiral}. We calculate the lattice momenta, wavefunction, and localization of the electrons on the zero-energy chiral channels. A topological explanation of the chiral modes is given in Section \ref{sec:top} including an argument for why the valley Chern number is a topological invariant of our system. In Section \ref{sec:strong}, we consider the regime where both the electric and magnetic fields are important. We numerically find how the energy bands, average position, localization, and layer preference depend on the magnetic and electric fields as well as the Fermi energy. Here, we introduce an approximation scheme to understand the prominent features in the case of large electric and magnetic fields. Concluding remarks are given in Section \ref{sec:con}.
 
\begin{figure}[b!] \label{fig:bilayerdiagram}
\centering
\includegraphics[width=0.45\textwidth]{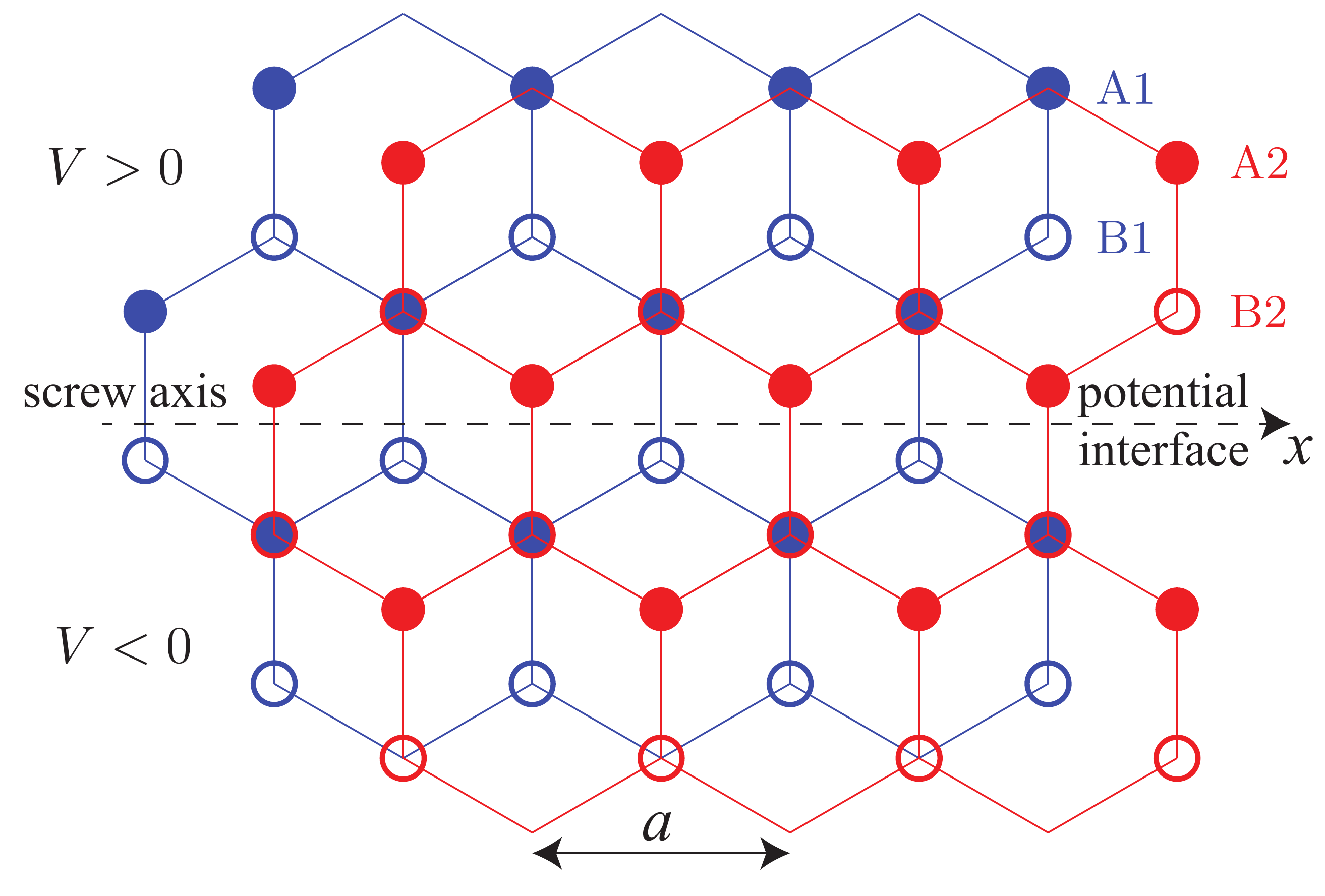}
\caption{AB-stacked bilayer graphene with electric domain wall at $y = 0$.} \label{blg_diagram}
\end{figure}

\section{Tight Binding Hamiltonian} \label{sec:tightbinding}

\begin{figure}
    \centering
    \raisebox{-.5\height}{\includegraphics[width=.12\textwidth]{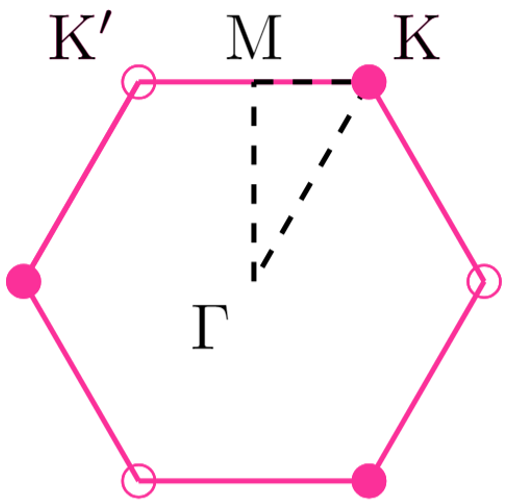}}
    \raisebox{-.5\height}{\includegraphics[width=.33\textwidth]{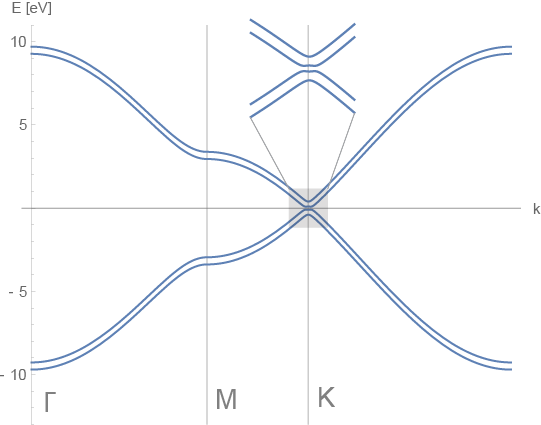}}    
    \caption{Band structure (right) of pristine bilayer graphene along a parameterized path through the $M$, $K$, and $\Gamma$ points in the Brillouin zone (left). The inset shows a gap at the $K$ point for $V = 0.1$ eV.}
    \label{fig:mkgamma}
\end{figure}

We review a tight-binding model Hamiltonian~\cite{McCann13} of bilayer graphene in the Bernal $AB$-stacking under a perpendicular electric field. Bilayer graphene may be described using a four atom unit cell consisting of the points $(A1, B1, A2, B2)$, describing the inequivalent $A$ and $B$ sites on each layer. The translations of this unit cell form a hexagonal Bravais lattice with the primitive vectors $\vb{a_{1}} = a(1,0)$ and $\vb{a_{2}} = a(-1/2, \sqrt{3}/2)$ with lattice constant $a = 2.46$ \AA\cite{Saito98}, as in monolayer graphene. AB-stacked bilayer graphene is formed by stacking two monolayer sheets with the two layers displaced by a distance $a/\sqrt{3}$. In our tight-binding approximation, we consider interactions between nearest neighbors. We ignore the spin degree of freedom due to weak spin-orbit interaction\cite{Yao07}, the direct $B1$ to $A2$ hopping, as well as next-nearest-neighbor and other weaker tunnelling processes. Using an intralayer nearest-neighbor hopping parameter $t = 3.16$ eV\cite{McCann13}, interlayer hopping parameter $u = 0.381$ eV between the $A1$ and $B2$ sites that lie directly on top of each other\cite{McCann13}, and electric potential energy difference $V$ between the layers, the tight-binding Hamiltonian can be written as
\begin{align}
    H = &- t\sum_{\ev{{\bf r},{\bf r}'}, s}(a_{{\bf r},s}^{\dagger}b_{{\bf r}',s} + \text{h.c.}) - u\sum_{{\bf r}}({a}_{{\bf r},1}^{\dagger}b_{{\bf r},2} + \text{h.c.}) \nonumber\\
    &-\frac{V}{2}\sum_{{\bf r},s}(-1)^s(a_{{\bf r},s}^{\dagger}a_{{\bf r},s} + b_{{\bf r},s}^{\dagger}b_{{\bf r},s}) \label{TBH}\\
		=&\int\frac{d^2{\bf k}}{(2\pi)^2}{\bf c}_{\bf k}^\dagger H({\bf k}){\bf c}_{\bf k}\nonumber
\end{align}
where ${\bf c}_{\bf k}=(a_{{\bf k},1},b_{{\bf k},1},a_{{\bf k},2},b_{{\bf k},2})^T$, $s=\{1,2\}$ sums over the layers, and ${\bf r}$ runs over Bravais lattice vectors $m_1{\bf a}_1+m_2{\bf a}_2$. The Bloch Hamiltonian is
\begin{equation} \label{eq:bloch}
H(\vec{k})=\begin{pmatrix}
V/2 & tf(\vec{k}) & 0 & u\\
tf^{*}(\vec{k}) & V/2 & 0 & 0\\
0 & 0 & -V/2 & tf(\vec{k})\\
u & 0 & tf^{*}(\vec{k}) & -V/2
\end{pmatrix}
\end{equation}
where $f(\vb{k}) = 1 + e^{-i\vb{k}\vdot\vb{a_{2}}} + e^{-i\vb{k}\vdot(\vb{a_{1}} + \vb{a_{2}})}$ are the interlayer hopping terms in momentum space describing the three nearest-neighbor points. Diagonalizing (\ref{eq:bloch}) gives a quadratic dispersion with zero energy gap when $V=0$ at the inequivalent $K$ and $K'$ points, on the edge of the Brillouin zone (Figure \ref{fig:mkgamma}). An energy gap $\Delta$ is introduced with a non-vanishing potential 
\begin{equation}
    \Delta = \frac{V}{\sqrt{1 + V^{2}/u^{2}}} \approx \begin{cases}V, & V \ll u \\ u, & V \gg u\end{cases}.
\end{equation}

If we are only concerned with low energy perturbations, we may expand $H$ about the $K$ and $K'$ points 
\begin{align}
H_{\mathrm{k}\cdot\mathrm{p}} &= [H_0({\bf k})+\nabla_{\bf k}H({\bf k})\vdot \delta {\bf k}]\big|_{{\bf k}=K,K'} \nonumber\\
&= \begin{pmatrix}
V/2 & \hbar v_F\pi & 0 & u\\
\hbar v_F\pi^{\dagger} & V/2 & 0 & 0\\
0 & 0 & -V/2 & \hbar v_F\pi \\
u & 0 & \hbar v_F\pi^{\dagger} & -V/2
\end{pmatrix}\label{eq:kp}\\
&=\hbar v_F(\nu\delta k_x\sigma_x-\delta k_y\sigma_y)\otimes\openone_2\nonumber\\
&\;\;\;+\frac{V}{2}\openone_2\otimes\tau_z+\frac{u}{2}(\sigma_x\otimes\tau_x-\sigma_y\otimes\tau_y)\nonumber
\end{align}
 where $\pi = \nu \delta k_{x} + i\delta k_{y}$, $v_F = \sqrt{3}at/2\hbar$ is the Fermi velocity for monolayer graphene, and $\nu$ is the valley index, $+1$ for $K$ and $-1$ for $K'$. Here, $\delta k_x$ and $\delta k_y$ are small momenta deviations away from the Dirac points, and the $2\times2$ Pauli matrices $\sigma$ ($\tau$) act on the $AB$-sublattice (layer) degree of freedom.

\section{Landau Levels} \label{sec:landau}
In this section, we discuss the other limit where the layer potential energy difference $V$ is absent and the bilayer system is under a uniform magnetic field $\vb{B} = B\vu{z}$. The momentum in the tight-binding Hamiltonian is replaced by the canonical momentum
\begin{equation}
    {\bf k}\longrightarrow {\bf k}-\frac{e{\bf A}}{\hbar}.
\end{equation}
The $\mathrm{k}\cdot\mathrm{p}$ Hamiltonian in \eqref{eq:kp} has the symmetry $[H_{\mathrm{k}\cdot\mathrm{p}}(K),Z]=[H_{\mathrm{k}\cdot\mathrm{p}}(K'),Z]=0$ for $Z=\sigma_z\otimes\tau_z$ at the Dirac points where $\delta{\bf k}=0$. This allows the partition of our basis into the dimer $(A2, B1)$ and $(A1, B2)$ sites that correspond to the two eigenspaces $Z=\pm1$. Focusing on one of the two eigenspaces, the energy spectrum near the Fermi level may be found by the effective two-band Hamiltonian\cite{McCann06,Pereira07} 
\begin{equation}
    H_{\mathrm{eff}} = -\frac{v_{F}^{2}}{u}\begin{pmatrix}0 & \pi^{2} \\ \qty(\pi^{\dagger})^{2} & 0\end{pmatrix}
\end{equation}
which is valid for $E \ll u, t$. We assume that $\psi_{A2}$ and $\psi_{B1}$ are shifted harmonic oscillator eigenstates. Through appropriate use of ladder operators, we may determine the energy spectrum
\begin{equation} \label{eq:landau}
    E_{n} = \operatorname{sgn}(n)\frac{2\hbar^{2}v_{F}^{2}}{ul_{B}^{2}}\sqrt{\abs{n}(\abs{n}-1)}
\end{equation}
where $l_{B} = \sqrt{\hbar/eB}$ is the magnetic length. 

The energy spectrum near the Dirac points is independent of momentum, corresponding to flat bands. The low-energy electrons in the bulk of the material are bound to cyclotron orbits. Choosing the Coulomb gauge $\vb{A} = -By\vu{x}$, our tight-binding Hamiltonian about the Dirac points is then given by (\ref{eq:kp}) with $V=0$, $\pi = \nu (\delta k_{x} + eBy/\hbar) +\partial_{y}$ and $\pi^{\dagger} = \nu (\delta k_{x} + eBy/\hbar) -\partial_{y}$, where we replace $k_{y} \to -i\partial_{y}$ because our Hamiltonian is now $y$-dependent. We seek zero-energy states in an infinite lattice; this requires two of the four components to be zero (which components are zero depends on the value of $\nu$). For concreteness, we consider the case $\nu=+1$. Solving for the two components introduces two arbitrary constants. One is fixed by enforcing normality, while the other is freely chosen to simplify calculations. We choose the remaining constant such that one of the solutions is purely localized to the top layer. The wavefunctions (suppressing the plane wave factor $e^{ik_xx}$) are
\begin{align}
    \psi_{B,1}(y) &= \frac{1}{\sqrt{\mathcal{N}}}e^{-(y+l_B^2 \delta k_x)^2/2l_{B}^{2}}
    \begin{pmatrix}0 \\ \sqrt{l_B^2/2+(\hbar v_F)^2/u^2} \\ 0 \\ 0 \end{pmatrix} \nonumber\\
    \psi_{B,2}(y) &= \frac{1}{\sqrt{\mathcal{N}}}e^{-(y+l_B^2 \delta k_x)^2/2l_{B}^{2}}
    \begin{pmatrix}0 \\ y+l_B^2 \delta k_x \\ 0 \\ -(\hbar v_F)/u \end{pmatrix}
\end{align}
where the normalization constant is
\begin{equation}
\begin{aligned}
		\mathcal{N} &= \frac{\sqrt{\pi} l_B^3}{2}\left[1+2\left(\frac{\hbar v_F}{u l_B}\right)^2\right]
\end{aligned}
\end{equation}
A general state at the lowest Landau level will then be a linear combination $\psi_{B} = \beta_{1}\psi_{B,1} + \beta_{2}\psi_{B,2}$. We observe that $\ev{y}=\bra{\psi_B} y \ket{\psi_B}$ depends linearly on the momentum $\delta k_{x}$ and is inversely proportional to the magnetic field strength. 
\begin{equation}
   \ev{y} = -l_{B}^{2} \delta k_{x}+\frac{l_B^2 u\Re(\beta_{1}^{*}\beta_{2})}{\sqrt{(\hbar v_F)^2+l_B^2 u^2/2}}
   \label{eq:yB}
\end{equation}
This result for $K'$ is similar in that $\ev{y}\propto -l_B^2 \delta k_x$. 

The localization length of $\psi_B$, $\sigma_y = \sqrt{\ev{ y^2} -\ev{y}^2}$, has no dependence on $\delta k_x$. 
\begin{equation}
    \sigma_y^2 = \frac{l_B^2 (\hbar v_F)^2+l_B^4 u^2[1+\Re(c_1^{*}c_2)-2\Im(c_1^{*}c_2)^2]}{2 (\hbar v_F)^2+l_B^2 u^2}
\end{equation}

\begin{figure}
    \centering
    \includegraphics[width=.45\textwidth]{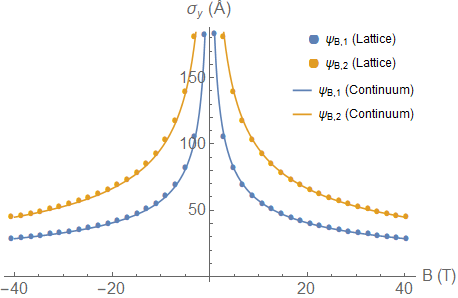}
    \caption{Localization of magnetic states. For large fields, the localization approaches $l_B/\sqrt{2}.$}
    \label{fig:stddev_purelyB}
\end{figure}

The asymmetry in the two nonzero components of $\psi_B$ suggests that in a magnetic field, electrons have a greater probability of occupying one of the two layers. Since we have two independent states, the layer preference will depend on how we take a linear combination of these two states. So, we can simply calculate the probability for an electron to occupy the bottom layer, $P_{\text{bottom}}$, by integrating the squared magnitude of the fourth component of $\psi_B$, which is just the fourth component of $\psi_{B,2}$
\begin{equation}
    P_{\text{bottom}} = \frac{1+|\beta_{2}|^2-|\beta_{1}|^2}{2+l_B^2 u^2/(\hbar v_F)^2}.
    \label{eq:pbot}
\end{equation}
$P_{\text{bottom}}$ is independent of $\delta k$, so for a fixed magnetic field and only small deviations away from the Dirac points, one can predict how the electrons localize to a certain layer. Even without knowledge of $\beta_1$ and $\beta_2$, one can put a bound on $P_{\text{bottom}}$
\begin{equation}
    0\leq P_{\text{bottom}} \leq \frac{2}{2+l_B^2 u^2/(\hbar v_F)^{2}}.
\end{equation}
Note that the lower and upper bounds are obtained using the states $\psi_{B,1}$ and $\psi_{B,2}$, respectively. For magnetic fields of up to 21 T, $l_B^2 u^2/(\hbar v_F)^2 > 10$, so the layer preference is approximately linear in $B$. This is the magnetic regime used in experimental setups~\cite{Li16}.
\begin{equation}
P_{\text{bottom}} = \frac{e \hbar v_F^2}{u^2} \qty(1+|\beta_{2}|^{2}-|\beta_{1}|^{2}) B.  
\end{equation}
At the $K'$ point, the opposite layer preference was found. 

\begin{figure}
    \centering
    \includegraphics[width=.5\textwidth]{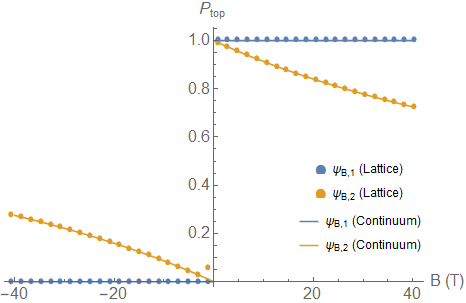}
    \caption{Probability for an electron to occupy the top layer. Any magnetic state that is a linear combination of these two states will have a similar curve that lies in between these two curves. For realistic fields, the layer preference is approximately linear in B.}
    \label{fig:prob_purelyB}
\end{figure}

\section{Chiral Modes} \label{sec:chiral}
Here, we summarize our findings from applying a first order $\mathrm{k}\cdot\mathrm{p}$ expansion about a Dirac point to bilayer graphene with a nonuniform electric field. Equation (\ref{eq:kp}) is modified by making the replacement $V\to V\operatorname{sgn}y$. This corresponds to introducing an electric field perpendicular to the lattice that switches direction at $y=0$ along the screw axis as shown in Figure \ref{blg_diagram}. We must again replace $\delta k_y \to -i\partial_{y}$ owing to translation symmetry breaking in $y$. Finding the eigenstates of this matrix requires solving a system of four coupled differential equations with appropriate boundary conditions at the interface. We explicitly found solutions in the zero-energy case. The solutions are proportional to $e^{i (k+ i \kappa) y}$ where $\kappa$ is given by
\begin{equation}
    \kappa = \frac{\abs{V}}{2\sqrt{2}\hbar v_f}\sqrt{-1+\sqrt{1+8\qty(\frac{u}{V})^2}}
\end{equation}
The value of $k$ is to be interpreted as the momentum where our chiral channels are located in reciprocal space.
\begin{equation}
    \label{eqn:dkx}
    k = \delta k_{x}=\frac{\abs{V}}{4\hbar v_F}\sqrt{1+\sqrt{1+8\qty(\frac{u}{V})^{2}}}.
\end{equation}
The solution takes the following form: 
\begin{equation}
\psi_{V} (y) = 
\begin{cases}
\psi_{V}^{-}(y), & y<0\\
\psi_{V}^{+}(y), & y\geq 0\\
\end{cases}
\end{equation}
\begin{equation}
\psi_{V}^{-}(y)=
\text{\footnotesize $
c_{V} \begin{pmatrix}
(V/2\hbar v_f) e^{i k y}+i (V/2\hbar v_f) e^{-i k y} \\ ((1-i)k-\kappa)e^{i k y}+((-1+i)k-i\kappa)e^{-i k y} \\ ((1-i)k-i \kappa)e^{i k y}+((-1+i)k-\kappa)e^{-i k y} \\ i (V/2\hbar v_f)e^{i k y}+(V/2\hbar v_f)e^{-i k y}
\end{pmatrix} e^{\kappa y}$}
\end{equation}
\begin{equation}
\psi_{V}^{+}(y)=
\text{\footnotesize $
c_{V} \begin{pmatrix}
(V/2\hbar v_f)e^{i k y}+i (V/2\hbar v_f) e^{-i k y} \\ ((-1+i)k-\kappa)e^{i k y}+((1-i)k-i \kappa) e^{-i k y}  \\ ((-1+i)k-i \kappa)e^{i k y}+((1-i)k-\kappa) e^{-i k y} \\ i (V/2\hbar v_f)e^{i k y}+(V/2\hbar v_f) e^{-i k y} 
\end{pmatrix} e^{-\kappa y}$}
\end{equation}
where the normalization constant is
\begin{equation}
c_{V}=\frac{\sqrt{\kappa}}{4 k}.
\label{eq:psiv}
\end{equation}
This describes a total of four zero-energy modes in the Brillouin zone, at $K\pm \delta k_x$ and $K' \pm \delta k_x$. The band structure indicates that two are right-moving modes near the $K$ point and the other two are left-moving near the $K'$ point that connect the valence and conducting bands. These chiral modes are exponentially localized directly on the electric interface, and the localization asymptotically approaches $\hbar v_F/u \approx 17.7$ \AA $\ $in the $V \gg 1$ eV limit. 
\begin{equation}
\sigma_y = \sqrt{\frac{1}{2\kappa^2}+\frac{\kappa^2}{2(k^2+\kappa^2)^2}}
\end{equation}

The probability distribution falls off as $e^{-y}$, unlike the magnetic states which fall off as $e^{-y^2}$. We provided further confirmation using numerical results from the tight-binding model.

\begin{figure}
    \centering
    \includegraphics[width=.45\textwidth]{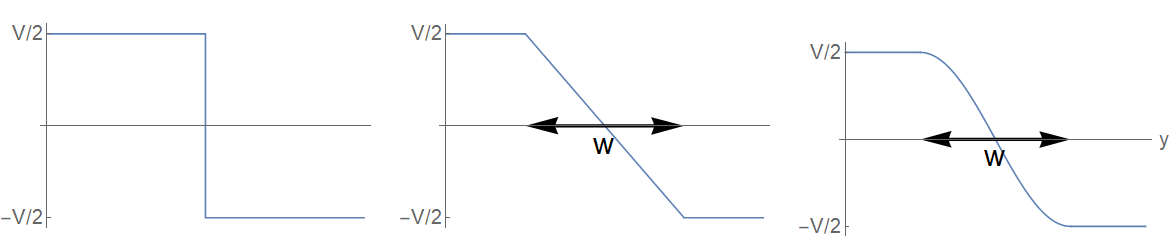}
    \caption{Three potential profiles used for calculations. The plots show the potential on the top layer; the bottom layer is given by the negative of this value. The linear and sinusoidal profiles are described by a width $w$, the characteristic gradient length scale.} \label{fig:potprofs}
    \includegraphics[trim={3cm 0 0 0},clip,width=.5\textwidth]{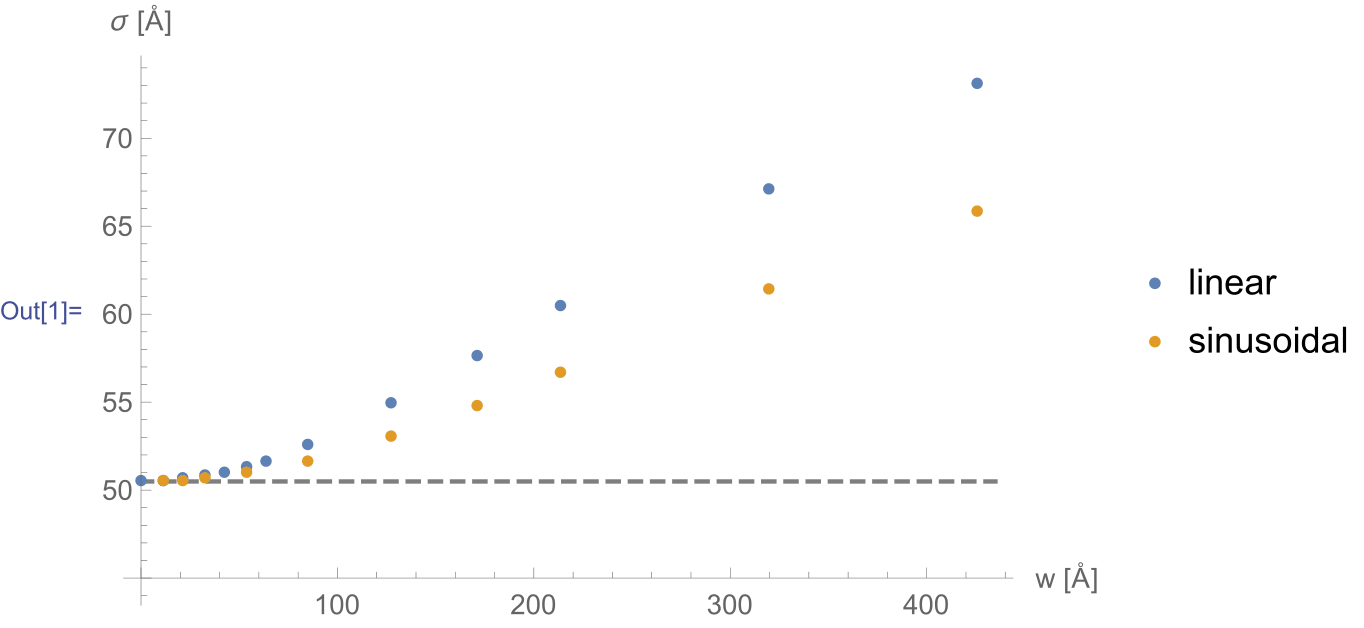}
    \caption{Localization of an interface state for different potential profiles as a function of $w$. Parameters used in the calculation are $E_{f}=0$ and $V=0.1$ eV. The dashed line indicates the step potential considered above.}
\end{figure}

We considered three different potential profiles: step-function, linear, and sinusoidal, as seen in Figure \ref{fig:potprofs}, the latter two having a characteristic gradient length scale $w$. All of these profiles produce two chiral channels near the $K$ point and two chiral channels near the $K'$ point. This is a result of the bulk topology, as will be described in the next section. The primary difference between the three profiles is the wavefunction localization; for a fixed $V$, the step potential leads to the most localized states. For simplicity, only the step function will be considered for the rest of the paper, though it is important to note that the localization can also be experimentally tuned by modulating the electric field. In the limit of small $w$ for a smoothly varying $V(y)$, one would then observe states with a localization approaching those in a step-function profile.

\begin{figure}
    \centering
    \includegraphics[width=.45\textwidth]{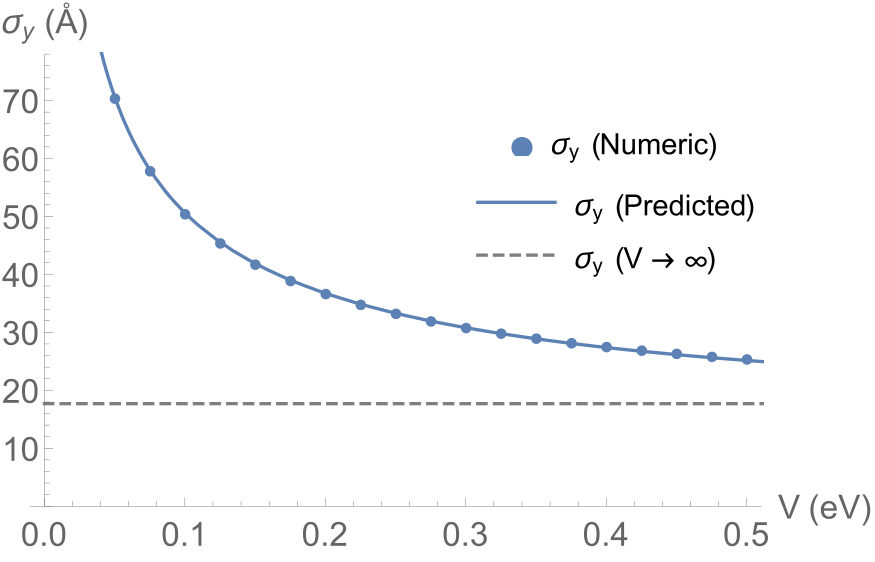}
    \caption{Localization of an interface state as a function of V. Plane wave solutions are recovered when the electric interface vanishes ($V=0$).}
    \label{fig:std_vs_V}
\end{figure}

\begin{figure}
    \centering
    \includegraphics[width=.45\textwidth]{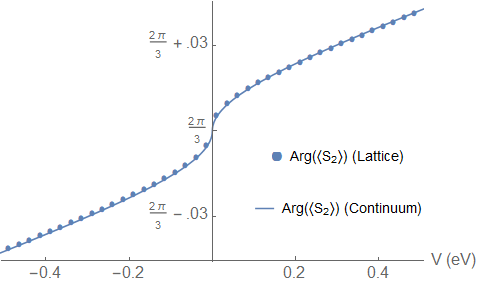}
    \caption{Argument of eigenvalues of screw operator at $K-\delta k_x$.}
    \label{fig:evals_screw}
\end{figure}

When a magnetic field is absent, bilayer graphene with the potential interface recovers a screw symmetry along the potential interface on the $x$-axis (see figure~\ref{blg_diagram}). Explicitly, we may write the symmetry transformation as $H(x, y, z, k_{x}) \to H(x+a/2, -y, -z, k_{x})$, where $z\to -z$ describes a layer interchange. The screw symmetry ensures equal probability distribution between the two layers. The screw symmetry is represented by a unitary matrix $S_2$ which squares to a unit translation in $x$, $S_2^2=e^{i{\bf k}\cdot{\bf a}_1}=e^{ik_xa}$. Therefore the screw eigenvalues are $\pm e^{ik_xa/2}$. Figure~\ref{fig:evals_screw} shows the effect of the potential on the argument of the screw eigenvalues for the zero-energy modes. In the $V=0$ limit, they begin at $\pm e^{i\pi/3}=\pm e^{iK\cdot{\bf a}_1/2}=\pm e^{iK_xa/2}$. Increasing or decreasing the potential shifts the Fermi momenta of the zero-modes to $K_x\pm\delta k_x$, where $\delta k_x$ takes the analytic form in \eqref{eqn:dkx} in the $\mathrm{k}\cdot\mathrm{p}$ approximation. The screw eigenvalues of the two zero modes become $\pm e^{i(\pi/3\pm\delta k_xa/2)}=\pm e^{i(K_x\pm\delta k_x)a/2}$. The evaluation of the argument of the screw eigenvalue $\langle\psi_0|S_2|\psi_0\rangle$ therefore provides a method to calculate the Fermi momentum shift $\delta k_x$ of the zero-mode $\psi_0$ in the discrete lattice tight-binding model. These values are shown as discrete points in Figure~\ref{fig:evals_screw}, and are well-approximated by the $\mathrm{k}\cdot\mathrm{p}$ approximation represented by the continuous curve.

\section{Bulk topology and symmetry considerations} \label{sec:top}
The existence of chiral modes is a consequence of the change of bulk topology across the electric interface. By the bulk-boundary correspondence~\cite{Volovikbook,Nakaharabook,RMP}, the number of gapless (valley) Dirac edge modes is identical to the change in (valley) Chern number\cite{TKNN,ZhangMacDonaldMele13} across the edge. The Chern number $\mathrm{Ch}_{1}$ is a topological property of the bulk, so by studying the topology of bilayer graphene with a perpendicular electric field, we can determine the behavior at the interface of these two configurations. $\mathrm{Ch}_1$ is typically calculated by integrating the Berry curvature $\vec{\Omega}$ over the Brillouin zone\cite{Tse11}
\begin{equation} \label{chernintegral}
    \mathrm{Ch}_1=\frac{1}{2\pi}\sum_n \int \dd^2 k (\vb{\Omega}_n)_z,
\end{equation}
where the sum is taken over the two occupied bands $n$ below the Fermi level at 0 energy and
\begin{equation} \label{berrycurvature}
   \vb{\Omega}_n=i \bra{\frac{\partial u_n}{\partial \vec{k}}} \cross \ket{\frac{\partial u_n}{\partial \vec{k}}}
\end{equation}
For bilayer graphene, the integral of the Berry curvature over the entire Brillouin zone vanishes as a result of time-reversal symmetry which requires the Berry curvature to be an odd function.\cite{Xiao10} However, the Berry curvature is localized near the $K/K'$ points, so one may consider the integral of the Berry curvature near these points. Using k$\cdot$p perturbation, we found the Berry curvature to be an odd function of $V$ and $\nu$. Integrating the local Berry curvature over all space gives the valley Chern number, though this is difficult to calculate using this formulation.


\begin{figure}[htbp]
    \centering
    \includegraphics[width=0.47\textwidth]{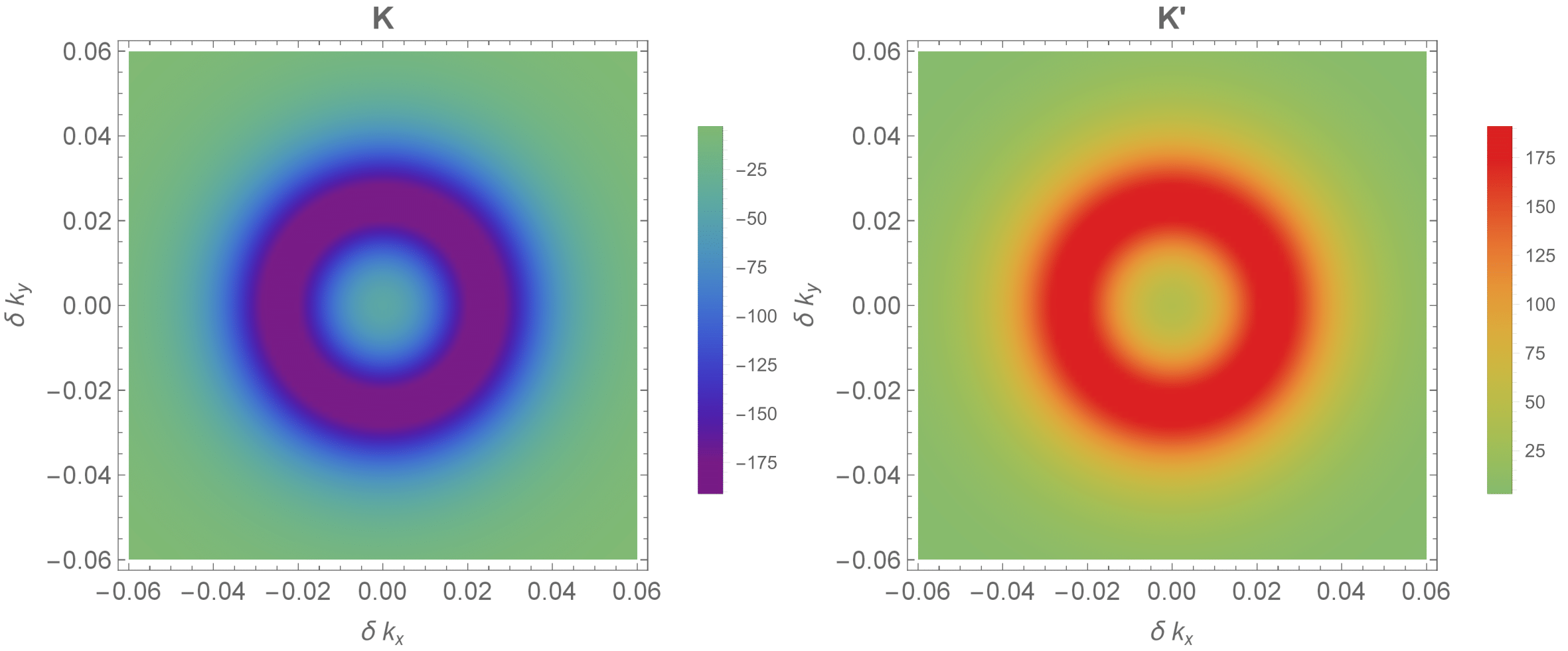}
    \caption{Berry curvature around the $K/K'$ point (left/right). At the $K'$ point, the sign of the Berry curvature is reversed, but the magnitude is identical.}
    \label{fig:berry_curvature}
\end{figure}

In the non-interacting limit, there is an equivalent formulation of the Chern number given in terms of the Green's function.\cite{MartinBlanterMorpurgo08, Wang12, Tse11, Volovikbook} Here, the Green's function takes on the form $G = (\tilde{E} - H(\vb{k}))^{-1}$
\begin{equation}
    \mathrm{Ch}_{1}=\frac{1}{24 \pi^2}\int\dd{\tilde{E}}\dd^2 k\, \mathrm{Tr}\bigg[\epsilon^{\mu \nu \rho} G \partial_{\mu}G^{-1} G\partial_{\nu} G^{-1} G\partial_{\rho} G^{-1}\bigg]
\end{equation}
where $\tilde{E}$ represents a complex energy and the energy integral is taken over a contour around the valence bands. The indices run from 0 to 2, where $\partial_{0}$ is an energy derivative and $\partial_{1}$ and $\partial_{2}$ are $k_x$ and $k_y$ derivatives, respectively. This expression is simpler to evaluate since it only involves derivatives of $\tilde{E}-H(\vec{k})$ rather than of the gauge-dependent normalized eigenstates as in \eqref{berrycurvature}. Since $\partial_0 G^{-1}=1$, we have that
\begin{equation}
    \mathrm{Ch}_{1}=\frac{1}{8 \pi^2}\int \dd{\tilde{E}} \dd^2 k\, \mathrm{Tr}\qty[G[G\partial_{1}G^{-1},G\partial_{2}G^{-1}]].
\end{equation}
The integrand only depends on the magnitude of the momentum, so the momentum integral can be evaluated in polar coordinates. The energy integral is found by evaluating the two residues at negative real energy to obtain
\begin{equation}
    \mathrm{Ch}_1=-\nu\operatorname{sgn} V.
\end{equation}
At the interface where $V$ changes sign, the difference in the Chern number is $-2/+2$ at the $K/K'$ point, in agreement with our previous observations of two left/right moving modes at the $K/K'$ point as obtained from the band structure.

\subsection{Symmetry protected topology}
The spinless bilayer graphene model \eqref{TBH} under a uniform electric field preserves time-reversal symmetry and charge $U(1)$ conservation, and thus belongs to class AI according to the tenfold classification~\cite{AltlandZirnbauer97} of band theory. In addition to these local symmetries, the model is also symmetric under the non-centrosymmetric space group (wallpaper group) P3m1, which is broken from the centrosymmetric P$\bar{3}$m1 by the layer-asymmetric electric potential. In particular, we focus on the point group symmetry $C_{3v}=\{1,r,r^2,\mu_x,r\mu_x,r^2\mu_x\}$, which is generated by a threefold rotations $r$ about the $z$-axis and a vertical mirror plane $\mu_x$ perpendicular to the $x$-axis. The twofold screw rotation described previously (in section~\ref{sec:chiral} and figure~\ref{fig:evals_screw}) only applies along the domain wall potential interface and is absent in the bulk where the electric field is uniform. Together with the time-reversal symmetry group $\mathbb{Z}_2^T$, which is generated by the anti-unitary time-reversal operator $T$ that squares to $T^2=+1$, they form the magnetic point group \begin{align}3m1'=\mathbb{Z}_2^T\times C_{3v}.\label{mpg}\end{align} In this subsection, we discuss the symmetry-protected topology.

\begin{figure}[htbp]
\centering\includegraphics[width=0.3\textwidth]{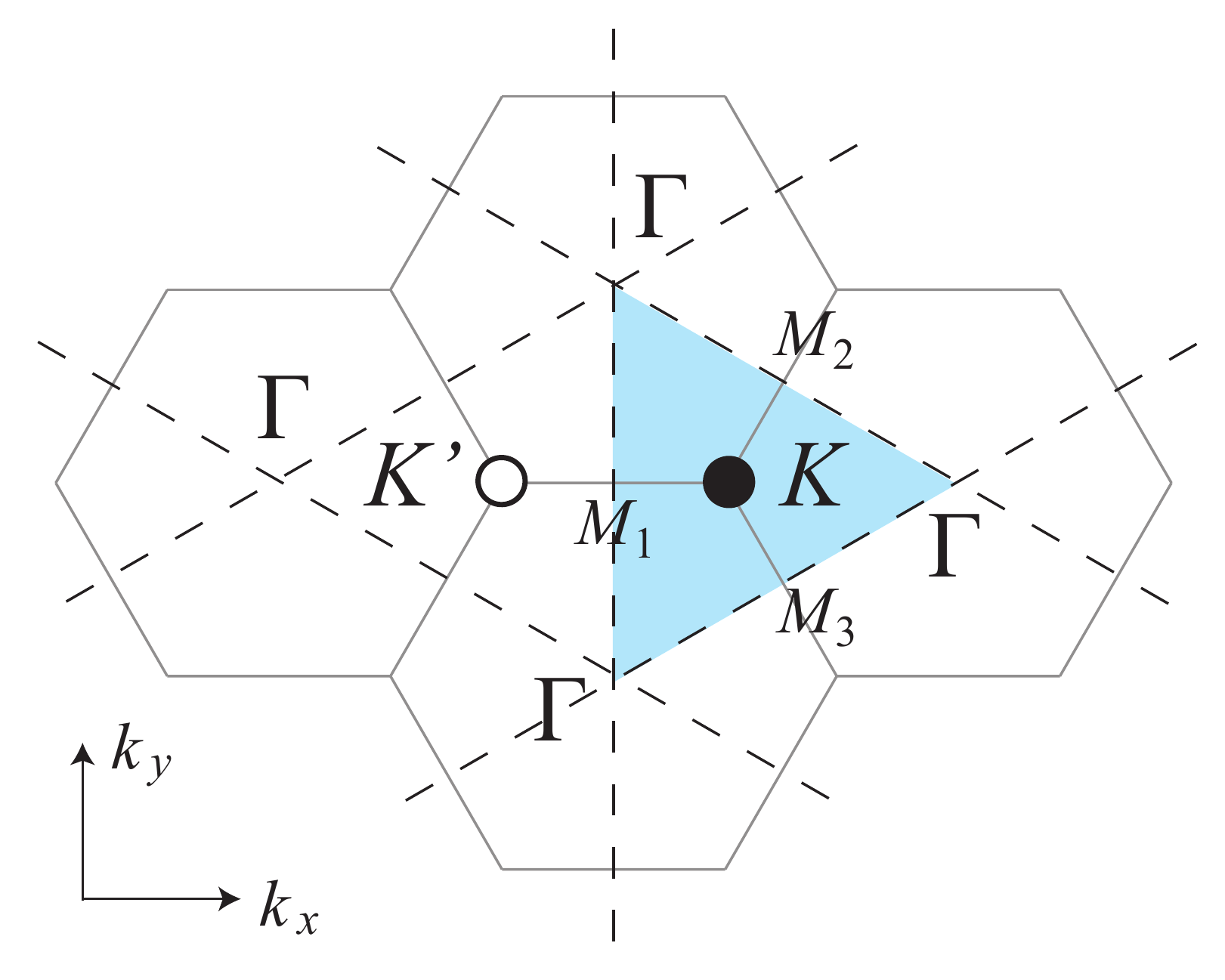}
\caption{The shaded triangular region represents the half Brillouin zone $\mathrm{BZ}^{1/2}_+$ that is used as the integration domain of the valley Chern number around $K$. The dashed lines represent the mirror-symmetric lines $L_1=\Gamma M_1\Gamma$, $L_2=\Gamma M_2\Gamma$ and $L_3=\Gamma M_3\Gamma$.}\label{fig:halfBZ}
\end{figure}

Mirror symmetry requires the Berry curvature to obey the asymmetry $\Omega_n({\bf k})=-\Omega_n(\mu_x{\bf k})$, where the minus sign is associated to the orientation reversing nature of reflection. Similar asymmetry relations hold for the other two mirror planes, $\Omega_n({\bf k})=-\Omega_n(r\mu_x{\bf k})=-\Omega_n(r^2\mu_x{\bf k})$. Therefore, as curvatures cancel between mirror opposite momentum points, the Chern number \eqref{chernintegral}, when integrated over the entire Brillouin zone, must be identically zero. In particular, along the three mirror-symmetric lines $L_1=\Gamma M_1\Gamma$, $L_2=\Gamma M_2\Gamma$ and $L_3=\Gamma M_3\Gamma$ where $\mu_x{\bf k}={\bf k}$, $r\mu_x{\bf k}={\bf k}$ and $r^2\mu_x{\bf k}={\bf k}$ respectively (see the dashed lines in figure~\ref{fig:halfBZ}), the Berry curvature must vanish, $\Omega_n|_{L_j}=0$. Here, $M_1,M_2,M_3$ are the three inequivalent time-reversal symmetric momenta apart from the Brillouin zone origin $\Gamma$. The three mirror-symmetric lines enclose a triangular region $\mathrm{BZ}^{1/2}_+$ (represented by the blue shaded region in figure~\ref{fig:halfBZ}) that contains a single $K$ point and traces out half of the Brillouin zone. Mirror (or time-reversal) maps $\mathrm{BZ}^{1/2}_+$ to another triangular region $\mathrm{BZ}^{1/2}_-$ that contains $K'$. Together, they generate the entire Brillouin zone $\mathrm{BZ}=\mathrm{BZ}^{1/2}_+\cup\mathrm{BZ}^{1/2}_-$. These two mirror-conjugated half-regions are almost mutually disjoint as they only intersects along the mirror-symmetric lines, $\mathrm{BZ}^{1/2}_+\cap\mathrm{BZ}^{1/2}_-=L_1\cup L_2\cup L_3$.

The valley Chern number is defined to be the integral \begin{align}v\mathrm{Ch}_1=\frac{1}{2\pi}\sum_{E_n({\bf k})<0}\int_{\mathrm{BZ}^{1/2}_+}d^2k\Omega_n({\bf k})\label{vChern}\end{align} over the half-Brillouin zone containing only a single valley. We now show that the point group symmetry $C_{3v}$ guarantees that the valley Chern number can only take discrete integral values. Consequently, $v\mathrm{Ch}_1$ is stable against any perturbation that preserves the $C_{3v}$ symmetries and excitation energy gap, and defines a symmetry-protected topological invariant. The Berry curvature is $d$-exact on the half-Brillouin zone, and is identical to the differential $\vb{\Omega}_n=\nabla_{\bf k}\times\boldsymbol{\alpha}_n$, where the Berry connection is \begin{align}\boldsymbol{\alpha}_n=i\left\langle u_n|\nabla_{\bf k}u_n\right\rangle.\end{align} From the Stokes' theorem, the valley Chern number is \begin{align}v\mathrm{Ch}_1=\frac{1}{2\pi}\sum_{E_n({\bf k})<0}\oint_{\partial\mathrm{BZ}^{1/2}_+}d{\bf k}\cdot\boldsymbol{\alpha}_n({\bf k}),\label{vChern2}\end{align} where the boundary of the half-Brillouin zone consists of the three mirror-symmetric lines, $\partial\mathrm{BZ}^{1/2}_+=L_1\cup L_2\cup L_3$. Without symmetries, the holonomy (also known as polarization) \begin{align}P_j=\frac{1}{2\pi}\oint_{L_j}d{\bf k}\cdot\boldsymbol{\alpha}_n({\bf k})\end{align} can take any real value. 
Time-reversal symmetry requires them to take integral values~\cite{FuKane06,QiHughesZhang08,RMP}. Each one of the mirror-symmetric lines $L_j$ is closed under time-reversal symmetry in the sense that if ${\bf k}$ belongs in $L_j$, so is its time-reversal conjugate $T{\bf k}=-{\bf k}$. Moreover, each line is a closed loop and is topologically equivalent to the 1D Brillouin zone. The band Hamiltonian $H({\bf k})|_{L_j}$ restricted on each of these lines is hence identical to a 1D time-reversal symmetric band insulator, which is known to have integral electric polarization. Combining the polarizations, the valley Chern number \eqref{vChern2} therefore has integral value. Lastly, since the Berry curvature is gauge invariant, so is the valley Chern number \eqref{vChern} and the sum of polarizations along $L_1$, $L_2$ and $L_3$ in \eqref{vChern2}.

We notice that the symmetry-protected topology actually only relies on the combination of mirror and time-reversal symmetry, rather than the individuals. These combinations form a symmetry subgroup \begin{align}3m'=\{1,r,r^2,T\mu_x,Tr\mu_x,Tr^2\mu_x\}\label{mpg0}\end{align} inside the full magnetic point group $3m1'$ in \eqref{mpg}. The valley Chern number must still take integral values based on this subset of magnetic point group symmetries. This is because the mirror-time-reversal combinations $T\mu_x,Tr\mu_x,Tr^2\mu_x$ take the same role as the local time-reversal $T$ along the mirror-symmetric lines $L_1,L_2,L_3$, and they still enforce the integrality of the polarizations $P_j$. In other words, the valley Chern number continues to provide a topological characterization of the system even when time-reversal symmetry is broken as long as the combined symmetries in \eqref{mpg0} are preserved. This applies, for instance, in the presence of a magnetic field on the bilayer graphene. This is because magnetic field is a pseudo-vector and is flipped under any improper rotation such as mirror. At the same time, it is also odd under time-reversal. Consequently, a magnetic field ${\bf B}=B_z\hat{\bf z}$ along the perpendicular direction is invariant under the magnetic point group $3m'$ in \eqref{mpg0}. Although some of the magnetic-mirror symmetries may be broken by the electric domain wall potential interface or a particular gauge choice ${\bf B}=\nabla\times{\bf A}$, we speculate that the robustness of the topological chiral channel along an electric domain wall in the presence of a magnetic field may be a consequence of the lingering non-trivial bulk topology protected by a magnetic symmetry. The numerical calculations presented in the following section will provide results consistent with this claim.

\section{Strong Electric and Magnetic Field} \label{sec:strong}
Tight-binding approximations were used to study bilayer graphene in the presence of strong electric and magnetic fields perpendicular to the bilayer in the lattice limit. The breaking of translation symmetry in $y$ with the introduction of an electric interface requires us to extend our unit cell to include the entire vertical dimension. To preserve translational symmetry in $x$, the Coulomb gauge $\vb{A} = -By\vu{x}$ is chosen with $y = 0$ set at the interface. 


Introducing a magnetic field requires the Peierls substitution in the hopping parameter
\begin{equation}
t\rightarrow t e^{i\theta}, \quad \theta\equiv\frac{2\pi}{\phi_0}\int_{r}^{r'}\vec{A}\vdot\dd{\vb{l}}
\label{eq:Peierls}
\end{equation}
where the integral is evaluated along the path between nearest-neighbor sites, and $\phi_{0} = h/e$ is the magnetic flux quantum. It will be useful to know when the magnetic energy scale is comparable to the electric energy scale set by $V$. The magnetic energy scale is set by the spacing between Landau levels. From equation (\ref{eq:landau}), $E_2-E_1=2(\hbar v_F)^2 e B \sqrt{2}/\hbar u\approx (0.005$ eV/T$)B$, so for $V\approx0.1$ eV, a comparable magnetic energy scale requires a field of about $B\approx20$ T. For many of the plots shown in this section, we use these values for $V$ and $B$. 

We numerically study this comparable limit in the tight binding model by considering lattice commensurate magnetic filling fractions, where the magnetic field takes on the form
\begin{equation}
    B = \frac{\phi_{0}}{\text{Area}}\frac{p}{q}
\end{equation}
where the rational number $p/q$ is the amount of magnetic flux (in unit of the flux quantum $\phi_0$) through a single hexagon plaquette. To study the topological chiral channels along the electric interface, we choose an open geometry along the vertical $y$ direction. Although a closed cylindrical geometry could avoid the irrelevant edge modes, the vector potential would become discontinuous because the geometry would enclose magnetic monopoles. The Peierls substitution \eqref{eq:Peierls} would in general be discontinuous as well unless the system circumference $L_y$ is also in some commensurate length. The discontinuity would correspond to an unphysical edge where the magnetic field diverges and additional edge modes arise. On the other hand, the Coulomb gauge vector potential is continuous in an open geometry. In a large system where the boundary edges at $y=L_y/2$ and $-L_y/2$ are far separated from the electric interface at $y=0$, the edge modes do not mix with interface states due to their exponentially localized wavefunction thanks to the bulk energy gap. For all of our calculations, we use at least $L_y=2000$ atoms in an open edge geometry. Although the edge modes are still visible in the electronic band structure, they are not of interest in our study. These irrelevant modes can be discarded by focusing only on quantum states that localized along the electric interface. Numerical analysis will be given for the relevant chiral modes near the $K$ point, and an explanation will be given on how to relate these results to the chiral modes near the $K'$ point.

\begin{figure}
    \centering
    \includegraphics[width=0.47\textwidth]{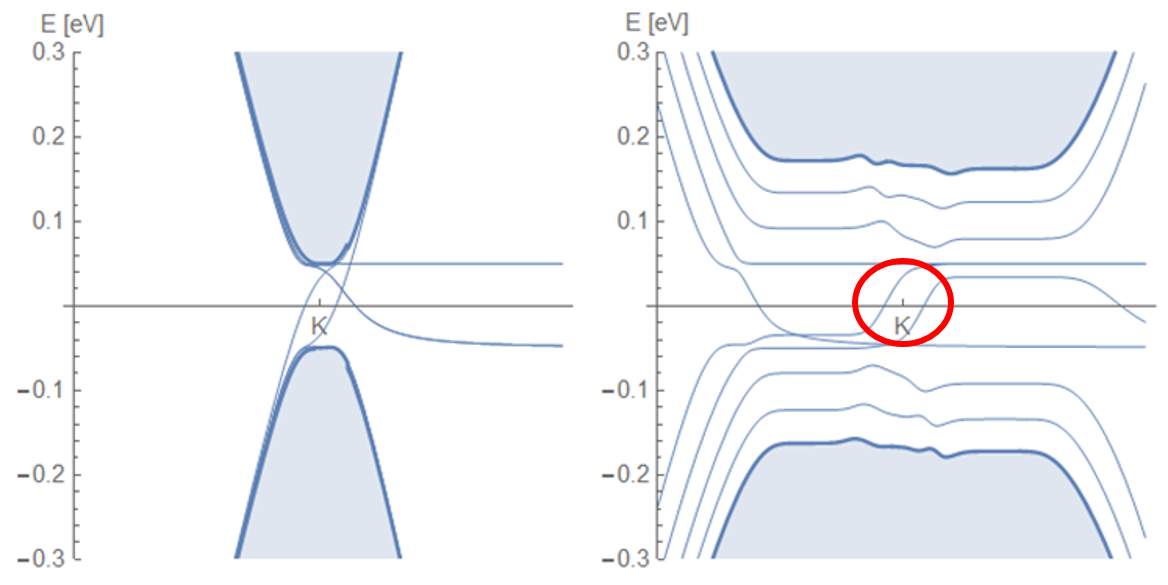}
    \caption{Electronic band structure with $B = 0$ T, $V = 0.1$ eV (left); $B = 20$ T, $V = 0.1$ eV (right), with $E_{f} = 0$ eV for both figures. The pairs of lines connecting the valence and conducting bands near the $K$ point are the chiral modes living on the interface (circled in the right figure), with a Fermi velocity given by the slope of the line. Additional lines crossing the Fermi level are localized states on the edge of the graphene strip, a result of using an open geometry in our calculations. Landau levels are observable on the right when $B \neq 0$.}
    \label{fig:bands}
\end{figure}

\begin{figure}
    \centering
    \includegraphics[width=.5\textwidth]{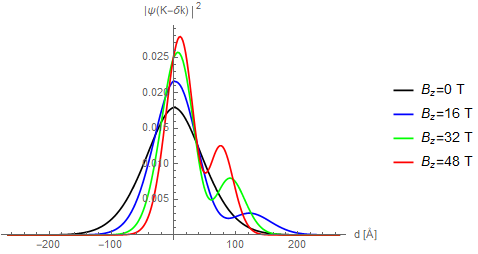}\hfill
    \includegraphics[width=.5\textwidth]{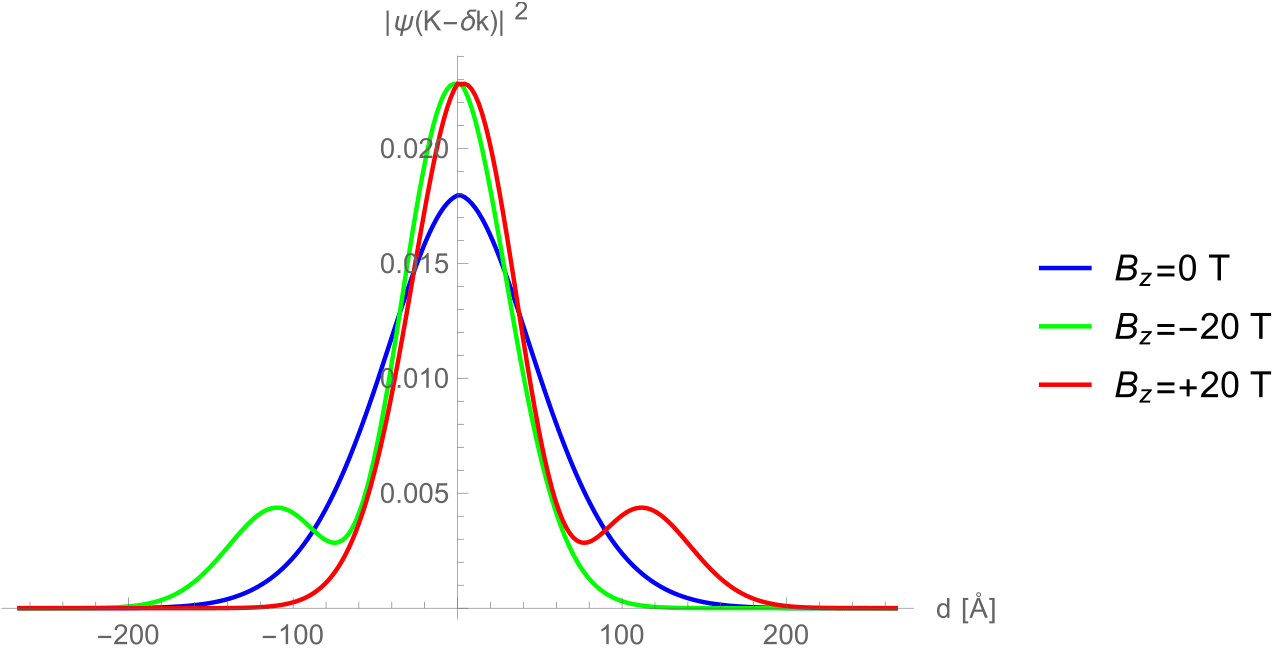}
    \caption{The shape of the wavefunction changes with the introduction of a magnetic field. The presence of a hump far from the interface becomes more prominent with a larger magnetic field. The side the hump forms on is determined by the sign of $B$. Here, $V=0.1$ eV and $E_f = 0$ eV.} \label{fig:ebstates}
\end{figure}

\begin{figure}
    \centering
    \includegraphics[width=.45\textwidth]{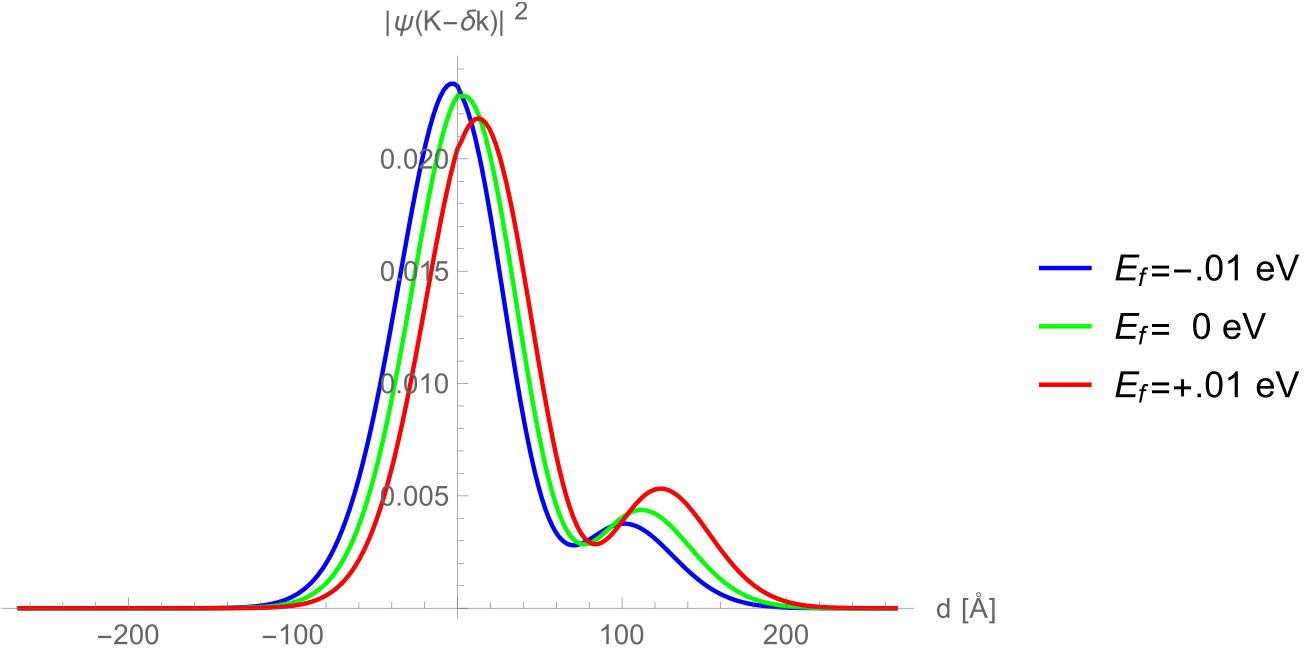}
    \caption{The peak of the wavefunction shifts with a nonzero Fermi energy when $B\neq0$. Here, $V=0.1$ eV and $B = 20$ T.} \label{fig:ebstates2}
\end{figure}

The band structure for bilayer graphene with an electric domain wall and no magnetic field indicates two right-moving modes near the $K$ point and two left-moving modes near the $K'$ point, as explained in section \ref{sec:chiral}. Plotting the wavefunctions corresponding to the chiral modes confirms that these states are exponentially localized at the domain wall as given by equation \ref{eq:psiv}. These wavefunctions have no dependence on the Fermi energy. The localization $\sigma_y$ in the numerical lattice tight binding model agrees with our continuum calculations, as shown in Figure \ref{fig:std_vs_V}. 

When both $B$ and $V$ are non-negligible, both flat bands (Landau levels) and interface modes can be observed in the band structure (Figure \ref{fig:bands}). Although classically one would expect electrons in a magnetic field to form cyclotron orbits leading to Landau levels at zero energy, the fixed chirality of the interface states prevents electrons from completing cyclotron orbits \cite{ZareniaPereiraFariasPeeters11}. The chiral modes are still localized near the interface, but their mean position $\ev{y}$ is not located exactly at the interface due to the formation of a secondary peak in the wavefunction (Figure \ref{fig:ebstates}). Electronic transport in this configuration would occur parallel to, but shifted away from, the electric domain wall. A nonzero Fermi energy also contributes to an overall shifting of the wavefunction peak (Figure \ref{fig:ebstates2}). This shifting only occurs when the magnetic field is nonzero.

Classically, one might expect the shifting of the edge modes from the electric domain wall to be related to how moving charges are bent in a magnetic field. By the Lorentz force, electrons on two different channels near $K$ moving in the same direction (or equivalently, electrons in the same valley) should be bent in the same direction, but we have found the co-propagating modes shift in opposite directions as the magnetic field is increased. The sign of the Fermi velocity is the same for the two modes at a fixed valley, but the force exerted on the electrons is opposite. Therefore, the Lorentz force does not provide an explanation of this phenomenon. We will later show that the shifting can be understood by the overlap of the electric and magnetic states at a fixed momentum.

\begin{figure}
    \centering
    \includegraphics[width=\linewidth]{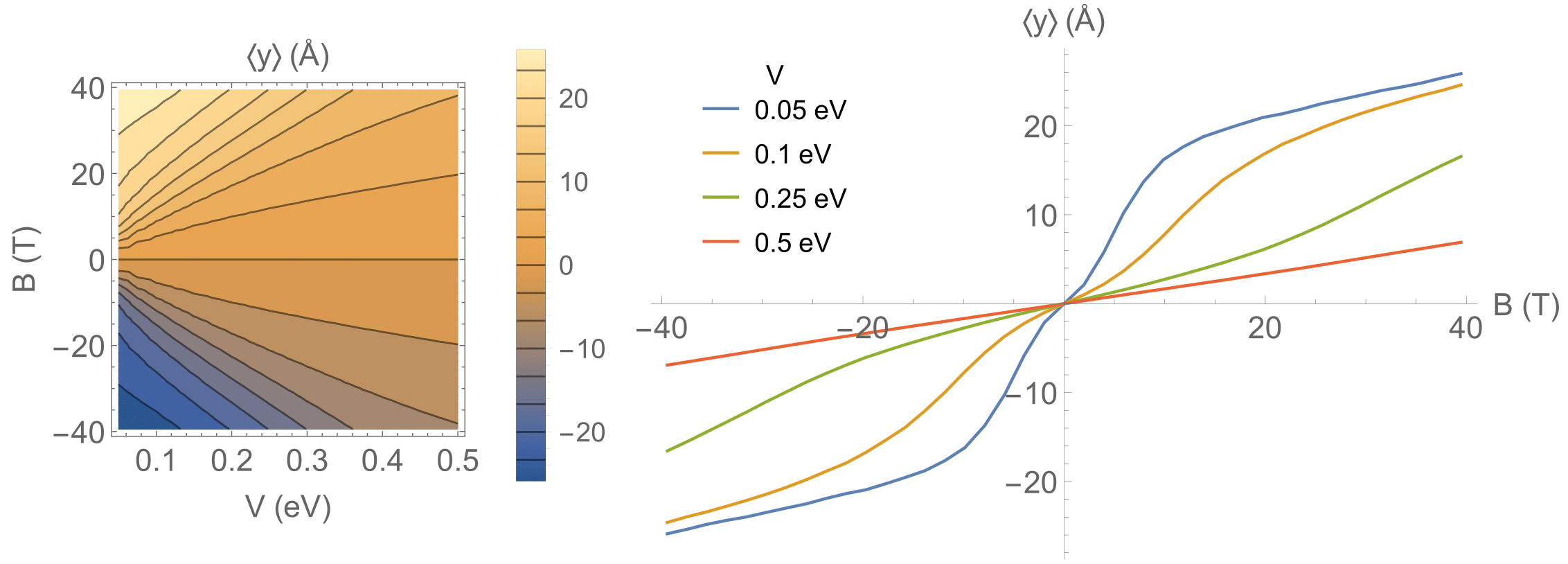}\hfill
    \caption{Overall shift of the wavefunction from the potential interface at $E_f = 0$.}
    \label{fig:vsBandVfixedEf}
\end{figure}

\begin{figure}
    \centering
    \includegraphics[width=\linewidth]{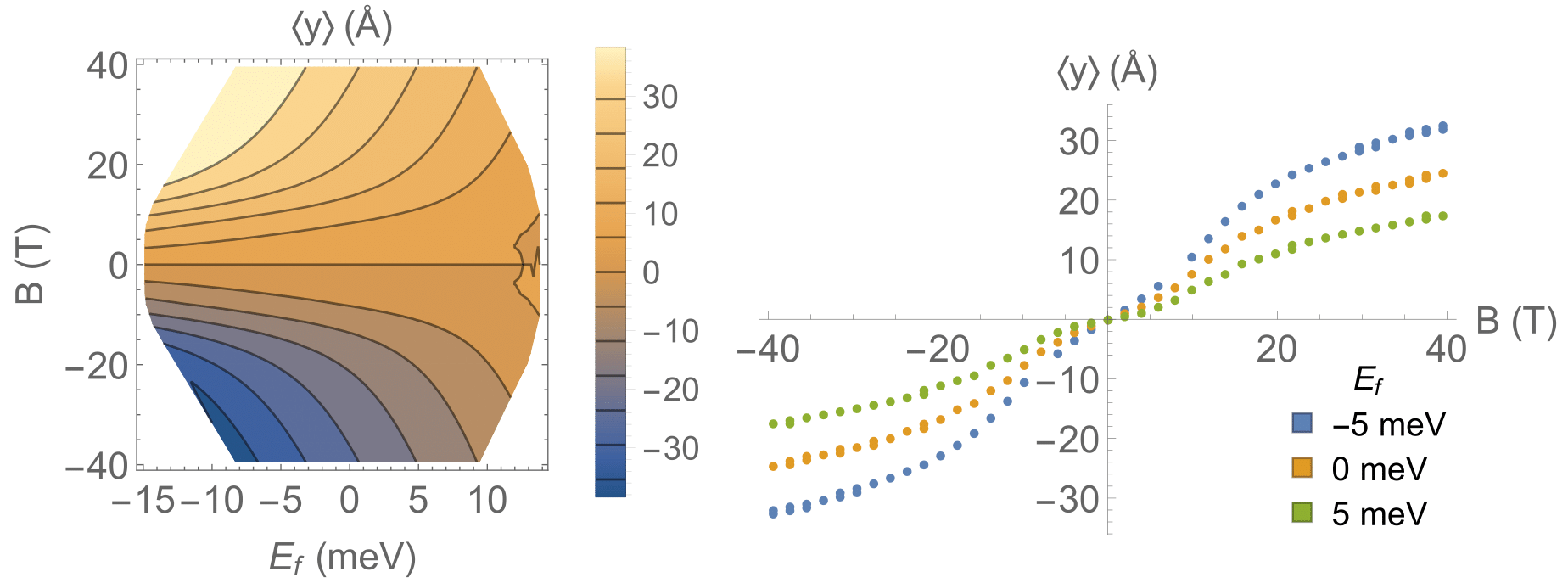}\hfill
    \caption{Overall shift of the wavefunction from the potential interface at $V=0.1$ eV.}
    \label{fig:vsBandEffixedV}
\end{figure}

\begin{figure}
    \centering
    \includegraphics[width=\linewidth]{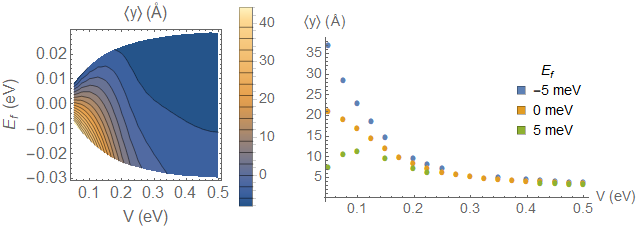}\hfill
    \caption{Overall shift of the wavefunction from the potential interface at $B=20$ T.}
    \label{fig:vsVandEffixedB}
\end{figure}

At a fixed Fermi energy, the deviation from the electric interface is directly proportional to the magnetic field strength with direction determined by the sign of B (Figure \ref{fig:vsBandVfixedEf}). Specifically, $\ev{y}(B)=-\ev{y}(-B)$. This shifting of $\ev{y}$ away from the interface is a result of the secondary peak in the wavefunction previously mentioned. Here, the electric field keeps the states localized to the interface while the magnetic field tends to shift the states away from the interface. Increasing the Fermi energy exaggerates the shifting for $B>0$ (Figure \ref{fig:vsBandEffixedV}). As one might expect, a small change in the Fermi energy does not have a significant effect on the shifting for larger $V$ because the Fermi level remains close to the center of the bulk band gap determined by $V$ (Figure \ref{fig:vsVandEffixedB}). 

\begin{figure}
    \centering
    \includegraphics[width=\linewidth]{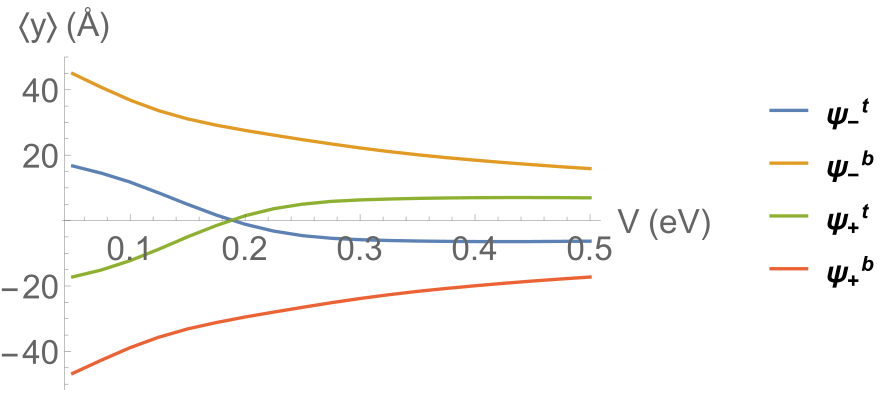}\hfill
    \caption{Overall shift of the wavefunction from the potential interface for $B=20$ T and $E_f=0$ eV. }
    \label{fig:vsVfixedBandEf}
\end{figure}

\begin{figure}
    \centering
    \includegraphics[width=\linewidth]{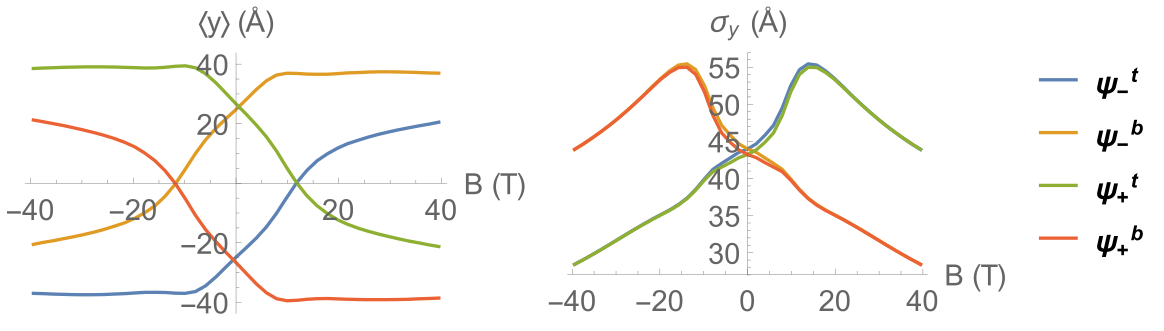}\hfill
    \caption{Overall shift and localization length of the wavefunction from the potential interface and standard deviation at $K$ for $E_{f} = 0$ eV and $V = 0.1$ eV.} 
    \label{fig:vsBfixedVandEf}
\end{figure}

The shifting can be better understood by partitioning the wavefunction into components residing on the top layer and those residing on the bottom layer. We use $\psi_-^t$ to denote the component of $\psi(K-\delta k_x)$ that lives on the top layer and $\psi_-^b$ for the bottom component of $\psi(K-\delta k_x)$. Similar notation is used for the top and bottom components of $\psi(K+\delta k_x)$. Note that while $\psi_{+}$ and $\psi_{-}$ shift in opposite directions, as $V$ becomes much larger than $B$, both wavefunctions approach the interface (Figure \ref{fig:vsVfixedBandEf}). Figure \ref{fig:vsBfixedVandEf} describes the shift and localization of the wavefunction as a function of $B$ for fixed $E_{f}$ and $V$. The peak in the figure on the right showing localization length $\sigma_y$ as a function of magnetic field is a result of two competing effects. For small magnetic fields, the magnetic contributions to the wavefunction are negligible, so $\sigma_y$ approaches that of the electric state. For very large magnetic fields, the magnetic states sit at the interface with a greater localization, consistent with equation \ref{eq:yB}. The peak occurs when the magnetic field strength is intermediate between these two scenarios. Then both the electric and magnetic states have non-negligible contributions, and the net effect is that the overall wavefunction broadens. The localization is identical for the four different chiral modes. 

\begin{figure}
    \centering
    \includegraphics[width=\linewidth]{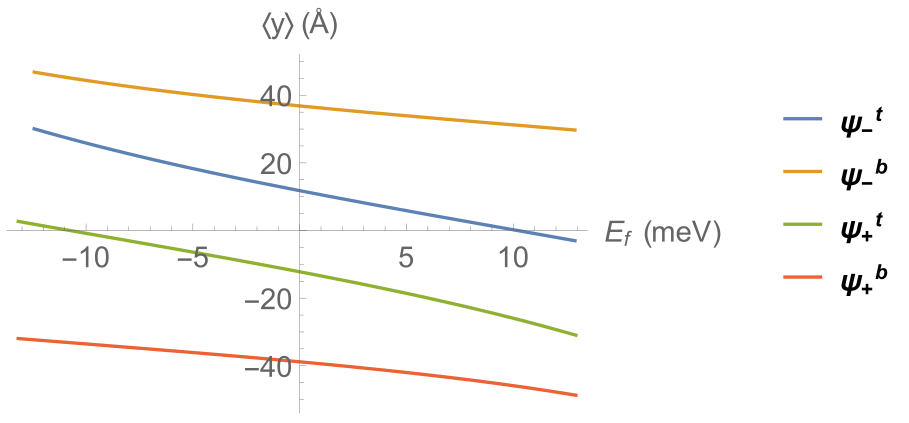}\hfill
    \caption{Layer shifting for the two states near the $K$ point. $B = 20$ T, $V = 0.1$ eV.}
    \label{fig:vsEffixedVandB}
\end{figure}

As the Fermi energy is increased, the two chiral modes in a given valley shift in the same direction (Figure \ref{fig:vsEffixedVandB}). The dependence of $\ev{y}$ on the Fermi energy is to be expected because $\ev{y}$ depends linearly on the Fermi momentum by equation \ref{eq:yB}. This differs from the $B=0$ case, where $\ev{y}=0$ regardless of the Fermi energy.

\begin{figure}
    \centering
    \includegraphics[width=.5\textwidth]{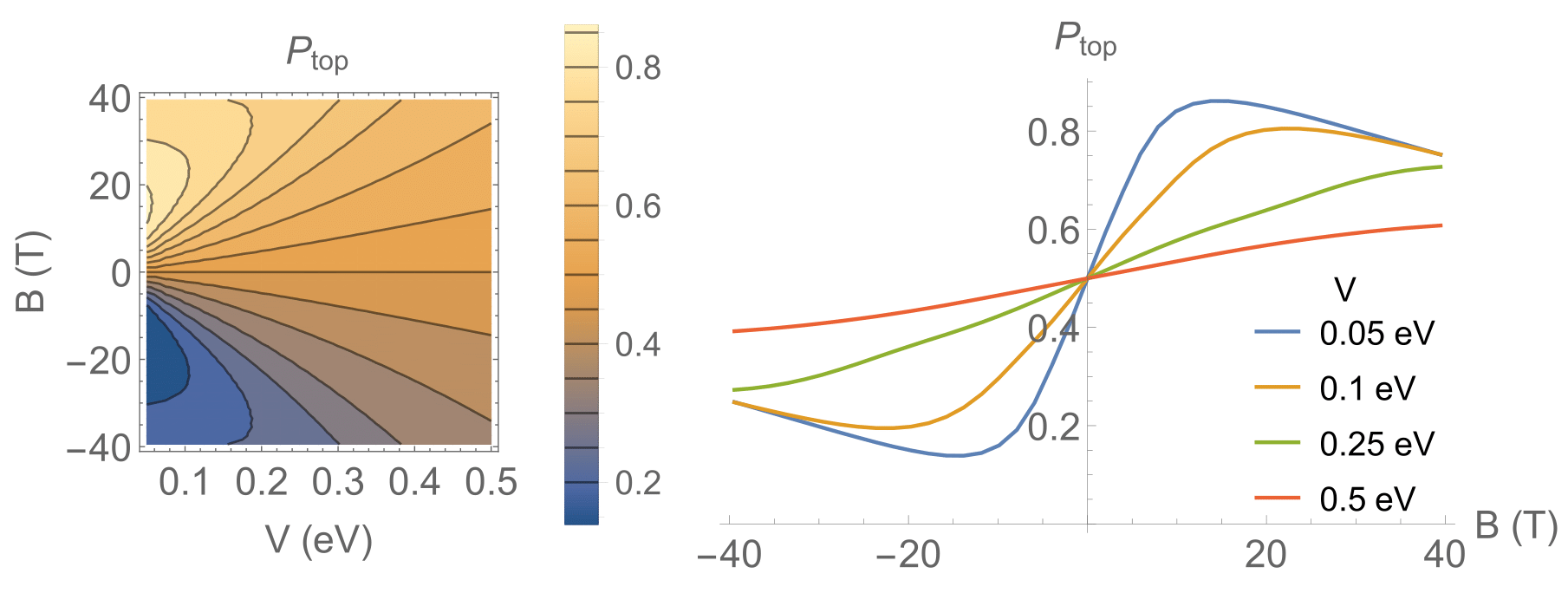}
    \caption{Probability for an electron to occupy the top layer.}
    \label{fig:layer_pref}
\end{figure}

As discussed before, bilayer graphene with $B = 0$, $V \neq 0$ obeys a screw symmetry along the axis of the interface. Since the vector potential is $y$-dependent, the symmetry is broken and a layer preference is allowed to form (Figure \ref{fig:layer_pref}). The layer preference depends on the valley index but not the sign of $\delta k_x$, consistent with the $V=0$ case. Furthermore, the layer preference has no dependence on the Fermi energy, since changing the Fermi energy only changes the Fermi momentum, and the layer preference does not depend on $\delta k_x$ as given in (\ref{eq:pbot}).

\begin{figure}
    \centering
    \includegraphics[width=.475\textwidth]{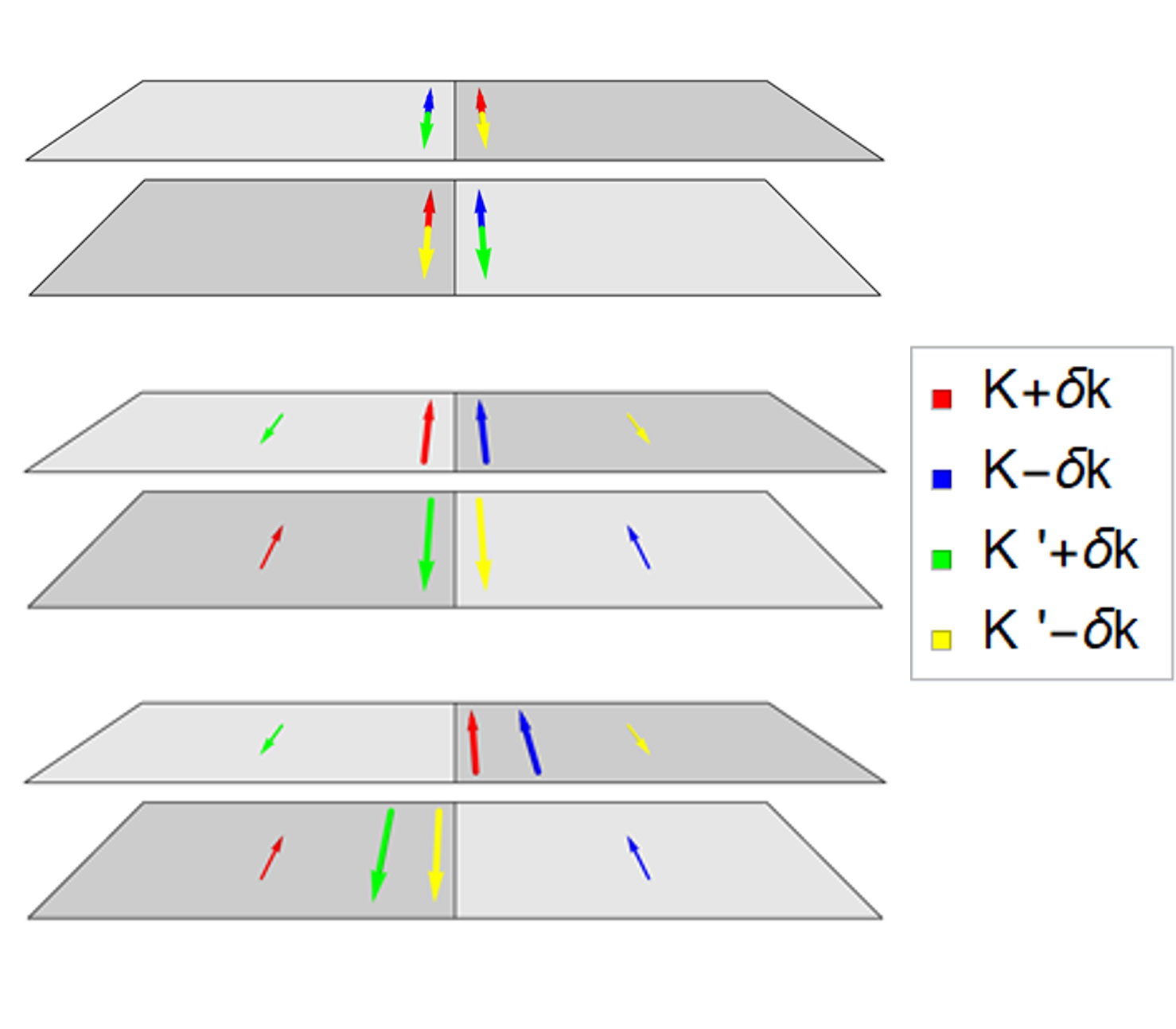}
    \caption{Electron propagation with $B=0,E_f=0$ (top), $B>0,E_f=0$ (middle), and $B>0,E_f>0$ (bottom) for electrons near the $K$ point. The size of the arrow is proportional to the probability of an electron to occupy that state.}
    \label{fig:electronpropagation}
\end{figure}

In order to understand why a magnetic field causes a peculiar shifting of the wavefunctions, we define $\psi_{BV}$ to be the linear combination of the electric and magnetic states
\begin{equation}
    \psi_{BV}(k_x)\approx \alpha \psi_{V}+\beta_1 \psi_{B,1}+\beta_2 \psi_{B,2}
\end{equation}
where $k_x = k_{x}(B, V)$ takes on the value of one of our four Fermi momenta and the coefficients $\alpha$, $\beta_{1}$, and $\beta_{2}$ are given by inverting the following matrix equation.
\small{
\begin{equation}
\begin{pmatrix}
1 & \bra{\psi_V}\ket{\psi_{B,1}} & \bra{\psi_V}\ket{\psi_{B,2}}\\
\bra{\psi_{B,1}}\ket{\psi_{V}} & 1 & 0\\
\bra{\psi_{B,2}}\ket{\psi_{V}} & 0 & 1
\end{pmatrix}
\begin{pmatrix}
\alpha \\ \beta_{1} \\ \beta_{2}
\end{pmatrix}
=
\begin{pmatrix}
\braket{\psi_V}{\psi} \\ \braket{\psi_{B,1}}{\psi} \\ \braket{\psi_{B,2}}{\psi}
\end{pmatrix}
\label{eq:bvapprox}
\end{equation}}

\begin{figure}
    \centering
    \includegraphics[width=\linewidth]{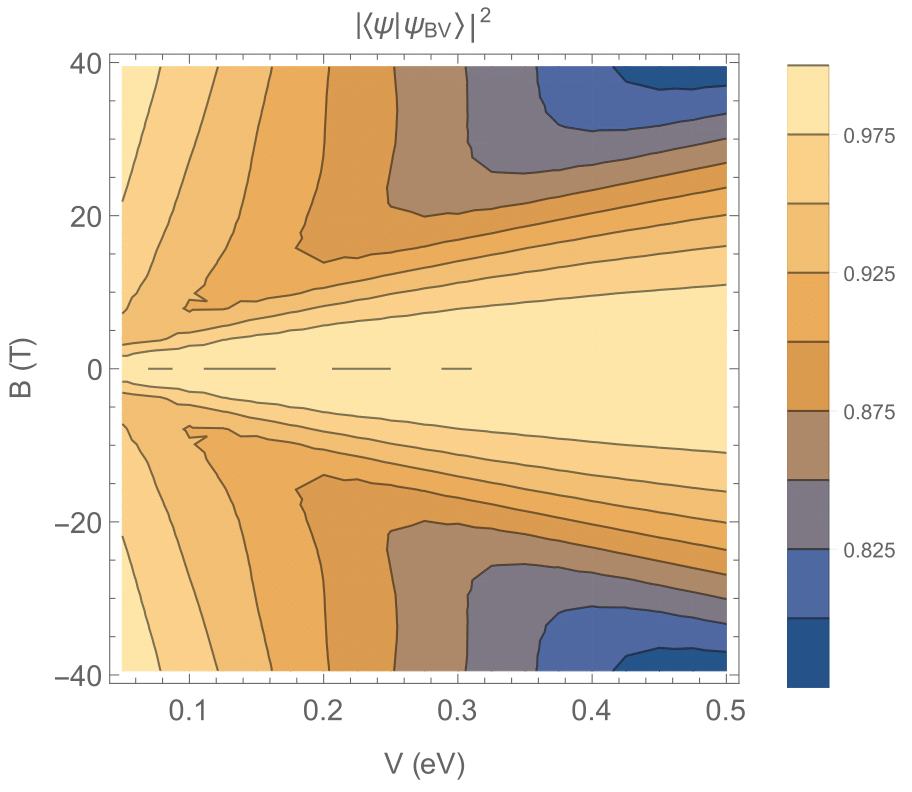}\hfill
    \caption{Modulus of the projection of our approximated state $\psi_{BV}$ onto the actual state $\psi$. This is a measure of the quality of our approximation.} \label{fig:approx}
\end{figure}

While the magnetic states are orthogonal to each other, the inner product between a magnetic state and the electric state generally does not vanish. Qualitatively, one might hope that $\psi_{BV}$ describes the actual wavefunction well. Our numerical results confirm that this is a good approximation for experimentally realizable fields (Figure \ref{fig:approx}). The magnitudes of $\alpha$ and $\beta$ give an indication of in what regimes $\psi_V$ dominates. 

Employing our approximation leads to a more intuitive understanding of the microscopic properties of the composite system. From our study of the magnetic states, we expect $\psi_B$ to shift in $y$ in proportion to $-\nu \delta k_x$. However, $\psi_V$ is localized at $y=0$ with momentum fixed by (\ref{eqn:dkx}). For this momentum, $\psi_B$ is localized at nonzero $y$. Therefore, the combination of these two separate wavefunctions with fixed momentum causes two peaks to form, the larger primary one directly on the interface is the contribution from $\psi_V$ and the smaller secondary hump shifted slightly from the interface is the contribution from $\psi_B$. It is important to stress that the hump \textit{approaches} the interface with increasing $B$ consistent with equation \ref{eq:yB}, but this causes a greater overlap with the electric and magnetic states, increasing the magnitude of the hump, and leading to an overall probability distribution shifted \textit{away} from the interface for experimentally realizable fields. In fact, for a small electric field $V \approx .1$ eV, the states will shift away from the interface with increasing $B$ for $B \lessapprox 50$ T. Figure \ref{fig:ebstates} displays this behavior. This explains the counter-intuitive notion that increasing the magnetic field shifts the magnetic wavefunction toward the interface, yet it shifts the overall wavefunction away from the interface for values of $V$ and $B$ used in experiment. In our approximation, this is equivalent to the more intuitive idea that strengthening $B$ increases the weight of our magnetic states in $\psi_{BV}$, as given by $|\beta|^2$.

\begin{figure}
    \centering
    \includegraphics[width=\linewidth]{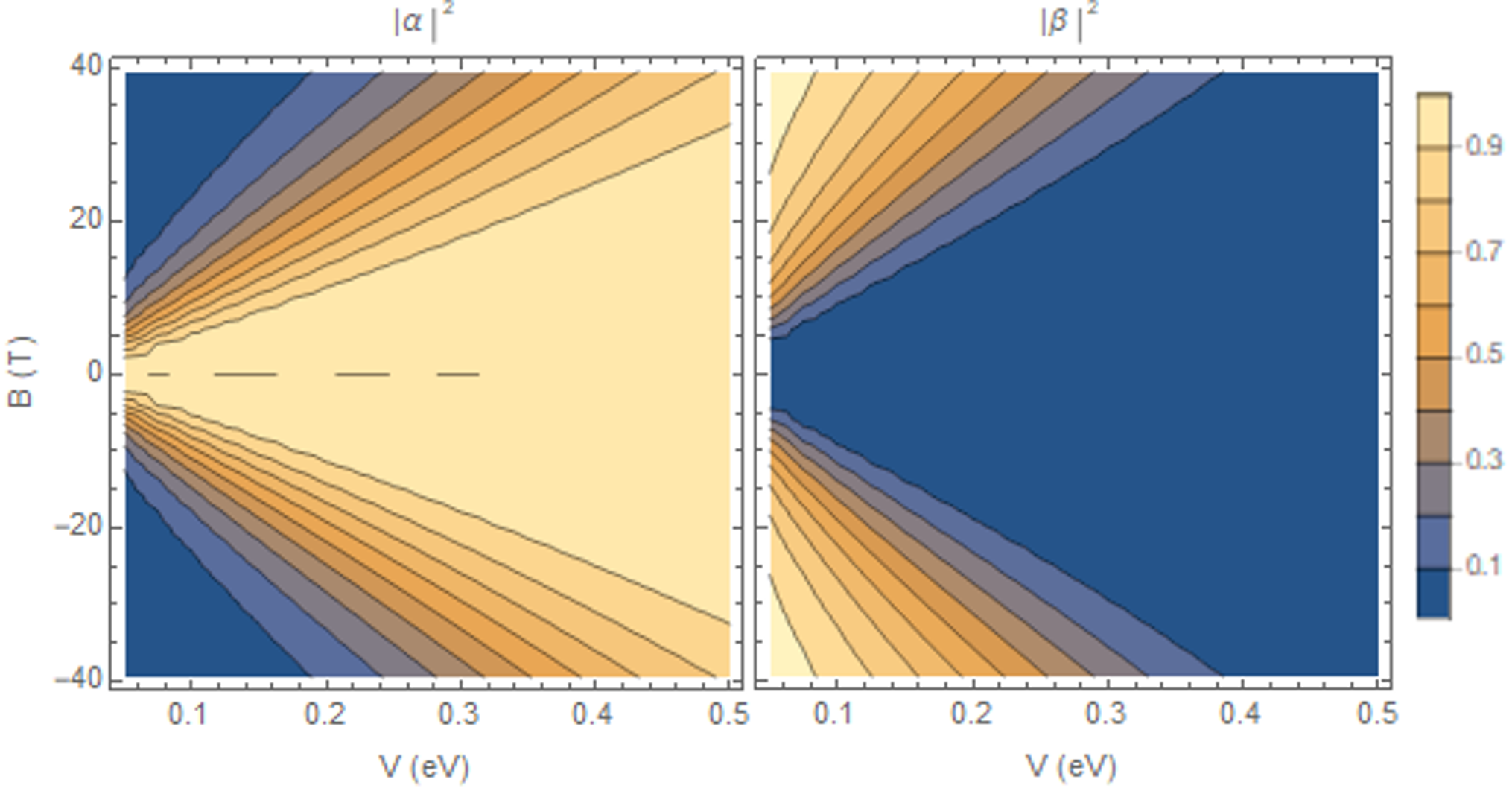}\hfill
    \caption{Modulus of the electric component $\alpha$ and magnetic component $\beta$ of $\psi_{BV}$, where $\beta$ is defined by $|\beta|^2=|\beta_1|^2+|\beta_2|^2$. The components change smoothly with varying $B$ and $V$.} 
\end{figure}

\section{Discussion and Conclusion} \label{sec:con}
It is well-known that the band gap of bilayer graphene can be tuned by changing the interlayer potential difference, such as by applying a perpendicular electric field. A perpendicular magnetic field provides an additional means of controlling the system. The magnetic field causes peculiar yet predictable alterations to the electronic wavefunction at electric domain walls. The lateral shifting and asymmetric layer distribution could partially explain the suppressed backscattering in the presence of a magnetic field as detected in experiments~\cite{Li16,Li18}. With no magnetic field, the counter-propagating modes at $K\pm \delta k$ and $K'\mp \delta k$ are located at the same position in real space. An introduction of a magnetic field separates these wavefunctions, decreasing their spatial overlap. For example, at $V=0.1$ eV and $E_f = 0$, $|\braket{\psi^{*}(K-\delta k)}{\psi(K'+\delta k)}|^2$ equals 1 when $B = 0$ but equals 0.73 when $B = 12$ T. While the counter-propagating modes at $K\pm \delta k$ and $K'\pm \delta k$ still lie at the same y-value, they are localized to opposite layers with a nonzero magnetic field.  This layer distribution is a new observation that could be detected using scanning tunneling microscopy.

The applications for valleytronic devices and transport measurements motivate a theoretical study for bilayer graphene with a spatially varying potential difference and a magnetic field. Attempting to diagonalize this Hamiltonian in the continuum limit results in a system of differential equations that admits no elementary solutions, its direct treatment being restricted to numerical methods. We present results showing that the actual solution can be approximated as a linear combination of the electric and magnetic states, all of which possess analytic forms. This combination of numerical and analytical results allows for high accuracy as well as a qualitative understanding of the microscopic properties. Unlike results obtained perturbatively, these results hold even in the regime that both $V$ and $B$ are large. Treating $\alpha$, $\beta_1$, and $\beta_2$ as pure numbers that can be found numerically by equation \ref{eq:bvapprox}, one may find equations for the average position, localization, and layer preference of the combined electric and magnetic state by using the results for the purely electric and purely magnetic states.  

\acknowledgments
We thank Prof.~Jun Zhu for insightful and inspiring discussion that initiated this work. We thank Matt Walker and Juan Velasco for helping with calculations and figures. We thank the Mead Endowment for supporting the undergraduate research program. JCYT is supported by the National Science Foundation under Grant No.~DMR-1653535.


%

\end{document}